--------------------------------------------------------------------

\documentstyle[12pt]{article}
\title{{\normalsize{{\hskip 8.5cm} BIHEP-TH-94-20}}\\[-7mm]
{\normalsize{{\hskip 8.5cm} June,~~~~~~1994}}\\
$q$-Deformed Chern Characters \\
for Quantum Groups $SU_{q}(N)$}

\author{  Bo-Yu Hou$^{1)}$, Bo-Yuan Hou$^{2)}$ and
Zhong-Qi Ma$^{3)}$\\
\parbox[t]{15cm}{{\footnotesize {1) Institute of Modern Physics,
Northwest University, Xi'an 710069, P. R. China }}\\[-2mm]
{\footnotesize {2) Graduate School, Chinese Academy of Sciences,
P. O. Box 3908, Beijing 100039, P. R. China}}\\[-2mm]
{\footnotesize {3) Institute of High Energy Physics, P. O. Box 918(4),
Beijing 100039, P. R. of China}}}}

\date{}
\begin{document}
\maketitle

\vspace{20mm}

\begin{abstract}
In this paper,  we introduce an $N\times N$ matrix $\epsilon^{a\bar{b}}$
in the quantum groups $SU_{q}(N)$ to transform the conjugate representation
into the standard form so that we are able to compute the explicit
forms of the important quantities in the bicovariant differential
calculus on $SU_{q}(N)$, such as the $q$-deformed structure constant
${\bf C}_{IJ}^{~K}$ and the $q$-deformed transposition operator
$\Lambda$. From the $q$-gauge covariant condition we define the
generalized $q$-deformed Killing form and the $m$-th $q$-deformed
Chern class $P_{m}$ for the quantum groups $SU_{q}(N)$. Some useful
relations of the generalized $q$-deformed Killing form are presented.
In terms of the $q$-deformed homotopy operator we are able to compute the
$q$-deformed Chern-Simons $Q_{2m-1}$ by the condition $dQ_{2m-1}=P_{m}$,
Furthermore, the $q$-deformed cocycle hierarchy,
the $q$-deformed gauge covariant Lagrangian, and the $q$-deformed
Yang-Mills equation are derived.

\end{abstract}

\newpage
\noindent
{\bf 1. INTRODUCTION}

Recently, quantum groups have attracted increasing attention. Since
the quantum group is provided by a noncommutative algebra, the
noncommutative geometry presented by Connes [Con] plays a basic
role like the differential geometry in the usual gauge theory.
Following the general ideas of Connes, Woronowicz [Wor1] [Wor2]
developed the framework of the noncommutative differential calculus.
He introduced the bimodule over the quantum group, and presented
various theorems concerning the differential forms and exterior
derivative. Manin [Manin] suggested a general construction of
quantum groups as linear transformations upon the quantum superplane.
The differential calculus on the quantum hyperplane was developed
by Wess and Zumino [WZ]. There have been a lot of papers treating the
differential calculus on quantum groups and the deformed gauge
theories from various viewpoints [Ber] [Jur] [Zum2] [CSWW] [IP]
[AC] [Cas] [SW] [FP] [Isa]. Here we would like to emphasize three
papers on the noncommutative differential geometry and deformation
of BRST algebra.

Aschieri and Castellani [AC] gave a pedagogical introduction to
the differential calculus on quantum groups by stressing at all
stages the connection with the classical case ($q~ \rightarrow ~1$).
As an example, they gave the explicit forms of some matrices appearing
in the bicovariant differential calculus on $SU_{q}(2)$. CSWW [CSWW]
presented a systematic construction of bicovariant bimodules on
the quantum groups $SU_{q}(N)$ and $SO_{q}(N)$ by using the $\hat{R}_{q}$
matrix. They described the conjugate of the fundamental representation
for $SU_{q}(N)$ as antisymmetrized product of $(N-1)$ fundamental
representations, and showed the expressions appearing in the
bicovariant differential calculus on $SU_{q}(N)$  both by formulas
and by diagrams. On the other hand, the antisymmetrized product
makes the calculation of explicit forms very complicated. In the later
paper [Wat] Watamura investigated the $q$-deformation of BRST algebra
for the quantum group $SU_{q}(2)$. Its generalization to $SU_{q}(N)$
depends on the explicit forms of the important quantities in
the bicovariant differential calculus on $SU_{q}(N)$, such as
the $q$-deformed structure constant ${\bf C}_{IJ}^{~K}$, the
$q$-deformed transposition operator $\Lambda$, and
projection operator ${\cal P}_{Adj}$.

Based on the $q$-deformed BRST algebra presented by Watamura,
we defined the $q$-deformed Killing form from the $q$-gauge
covariant condition, and construct the second $q$-deformed Chern
class, $q$-deformed Chern-Simons, and the cocycle hierarchy for
$SU_{q}(2)$ in our previous paper [HHM]. In order to investigate
the $SU_{q}(N)$ gauge theory we have to compute the explicit forms
of the quantities in the bicovariant differential calculus on
$SU_{q}(N)$.

In fact, the key for solving this problem is to change the
description for the conjugate representation. As everyone
knows, the conjugate of the fundamental representation in $SU(N)$
is equivalent to the antisymmetrized product of $(N-1)$ fundamental
representations, as used by [CSWW]. However, it is also equivalent to
a basic highest weight representation described by the last fundamental
dominant weight ${\bf \lambda}_{N-1}$. The monoid $\epsilon^{a\bar{b}}$,
that is an $N \times N$ matrix, plays a very important role in the
explicit calculations. The monoid $\epsilon^{a\bar{b}}$, that seems
to have some relation with the $q$-deformed Weyl element [KR], is
proportional to the $q$-deformed Clebsch-Gordan matrix reducing the direct
representation space of ${\bf \lambda}_{1} \otimes {\bf \lambda}_{N-1}$
into the identity representation space [Ma], and provides the relations
among the relevant $\hat{R}_{q}$ matrices. On the other hand, it
also serves as the similarity transformation from the conjugate
of the fundamental representation to the highest weight representation
${\bf \lambda}_{N-1}$. From those $\hat{R}_{q}$ matrices and
monoid we compute the explicit forms of the quantities appearing
in the bicovariant differential calculus on $SU_{q}(N)$ so that
we are able to generalize the $q$-deformed $SU_{q}(2)$ gauge theory
to the quantum groups $SU_{q}(N)$. In other words, for the quantum
groups $SU_{q}(N)$, we generalize the $q$-deformed BRST algebra,
define the $q$-deformed Killing form, and construct the $m$-th
$q$-deformed Chern class and $q$-deformed cocycle hierarchy. In [CSWW] and
[CW] a similar matrix $\epsilon^{i[j]}$ was introduced, where $[j]$
is the short form for $(N-1)$ antisymmetrized indices. However, our
$\epsilon^{a\bar{b}}$ makes the calculations much simpler.

Recently, Isaev [Isa] discussed the $q$-deformed Chern characters
where the base manifold is the $q$-deformed coset space
$\left(GL_{q}(N+1)/(GL_{q}(N)\otimes GL(1)\right)$.
In the present paper, just like in [CSWW] and [AC], the spacetime is
taken to be the ordinary commutative Minkowski spacetime, while the
$q$-structure resides on the fiber, the gauge potentials being
non-commutating. In the theory there are two nilpotent operators:
the BRST transformation $\delta$ and the derivative $d$, such that
we can discuss the double cohomology and cocycle hierarchy.
In terms of a $q$-deformed homotopy operator we are able to compute
the $q$-deformed Chern-Simons $Q_{2m-1}$ by the condition $dQ_{2m-1}=P_{m}$.
As example, we write the explicit forms of the
$q$-deformed Chern-Simons $Q_{3}$ and $Q_{5}$ explicitly.

The plan of this paper is as follows. In Sec. 2 we calculate the
$\hat{R}_{q}$ matrices in the product representation spaces of
${\bf \lambda}_{1}\otimes {\bf \lambda}_{1}$,
${\bf \lambda}_{N-1}\otimes {\bf \lambda}_{N-1}$,
${\bf \lambda}_{1}\otimes {\bf \lambda}_{N-1}$, and
${\bf \lambda}_{N-1}\otimes {\bf \lambda}_{1}$, and discuss
their main properties. The
monoid $\epsilon^{a\bar{b}}$ is introduced to relate those
$\hat{R}_{q}$ matrices. The algebra of functions on the quantum
group $SU_{q}(N)$ is sketched in Sec. 3. It is proved that the
monoid transforms the conjugate of the fundamental representation
${\bf \lambda}_{1}$ into the highest weight representation
${\bf \lambda}_{N-1}$. The generalized $q$-Pauli matrices are
defined to separate the singlet from the adjoint components.
In Sec. 4 we review the bicovariant differential calculus on
$SU_{q}(N)$, and compute the explicit forms of some important
quantities. Generalizing Watamura's investigation, the $q$-deformed
BRST algebra for the quantum group $SU_{q}(N)$ is constructed in
Sec. 5. From the condition $\delta P_{m}=0$ and $dP_{m}=0$, we define
the $q$-deformed generalized Killing form and the $m$-th $q$-deformed Chern
class $P_{m}$ in Sec. 6. In Sec.7, the $q$-deformed homotopy operator
is introduced in $SU_{q}(N)$ to compute the $q$-deformed
Chern-Simons. Furthermore, the $q$-deformed
cocycle hierarchy and $q$-deformed Yang-Mills equation for $SU_{q}(N)$
gauge theory are obtained in Sec. 7 and Sec. 8. Just like in the case
of $SU_{q}(2)$, the components of the identity and the adjoint
representations are also separated in the $q$-deformed $SU_{q}(N)$
gauge theory, although they are mixed in the commutative relations
of BRST algebra.

\vspace{10mm}
\noindent
{\bf 2. $\hat{R}_{q}$ Matrices and Monoid}

In the quantum enveloping algebra $U_{q}A_{N-1}$ there are $(N-1)$
simple roots ${\bf r}_{n}$ and $(N-1)$ fundamental dominant weight
${\bf \lambda}_{m}$, that are related by the Cartan matrix $a_{mn}$:
$$\begin{array}{l}
{\bf r}_{n}~=~{\bf \lambda}_{m}~a_{mn},~~~~n,~m=1,~2,\cdots, (N-1) \\
a_{mn}~=~d_{m}^{-1}~\left({\bf r}_{m}\cdot {\bf r}_{n}\right),~~~~
d_{m}~=~\left({\bf r}_{m}\cdot {\bf r}_{m}\right)/2
 \end{array} \eqno (2.1) $$

\noindent
In the present paper, if without special notification, summation
of the repeated indices is understood. To make formulas more
symmetrical, we define:
$${\bf \lambda}_{0}~=~{\bf \lambda}_{N}~=~0 \eqno (2.2) $$

A highest weight representation is denoted by its highest weight,
that is a positive integral combination of ${\bf \lambda}_{j}$.
The states in the representation are described by their weights
that are the integral combinations of ${\bf \lambda}_{j}$.
In this paper we are interested in only eight representations: the
basic representations ${\bf \lambda}_{1}$, ${\bf \lambda}_{2}$,
${\bf \lambda}_{N-2}$ and ${\bf \lambda}_{N-1}$, the symmetrical
tensor representations $2{\bf \lambda}_{1}$ and $2{\bf \lambda}_{N-1}$,
the adjoint representation ${\bf \lambda}_{1}+{\bf \lambda}_{N-1}$,
and the identity representation with the highest weight ${\bf 0}$.
Sometimes, the representations ${\bf \lambda}_{2}$ and ${\bf \lambda}_{N-2}$
are also called the antisymmetrical tensor representations.

The fundamental representation ${\bf \lambda}_{1}$ is $N-$dimensional.
The states in this representation are described by their weight
$({\bf \lambda}_{a}-{\bf \lambda}_{a-1})$. For simplicity, we
enumerate the states by one index $a$ as usual:
$$a~\rightarrow~ {\bf \lambda}_{a}-{\bf \lambda}_{a-1},~~~~
a=1, ~2,\cdots, ~N \eqno (2.3)$$

\noindent
The conjugate of the representation ${\bf \lambda}_{1}$ is equivalent
to the representation ${\bf \lambda}_{N-1}$, where the states have
the weights ${\bf \lambda}_{a-1}-{\bf \lambda}_{a}$ and are enumerated
by one index $\bar{a}$:
$$\bar{a}~\equiv~-~a~\rightarrow~ {\bf \lambda}_{a-1}-{\bf \lambda}_{a},~~~~
\bar{a}=\bar{N}, \cdots,~\bar{2}, ~\bar{1}, \eqno (2.4)$$

\noindent
In the present paper, if without special notification, the small Latin letter
except for $n$ and $m$, such as $a$ and $i$, runs over $1,~2,~\cdots, ~N$.

The standard method for calculating the solutions $\hat{R}_{q}$ of the
simple Yang-Baxter equation [Res] [Ma] is to expand it by the
projection operators that are the products of two quantum Clebsch-Gordan
matrices. The calculation method for $\hat{R}_{q}$ was described in the
textbook [Ma] in detail. In order to fit the usual notation in the theory
of the quantum groups, the solution $\hat{R}_{q}$ here is related to the
solution $\breve{R}_{q}$ used in the book [Ma] as follows:
$$\hat{R}_{q}~=~q~\breve{R}_{q}^{-1} \eqno (2.5) $$

In the direct product space of ${\bf \lambda}_{1}\otimes {\bf
\lambda}_{1}$ the solution of the simple Yang-Baxter equation  is:
$$\left(\hat{R}_{q}^{{\bf \lambda}_{1}{\bf \lambda}_{1}}\right)^{ab}_{~cd}~=~
q~\left({\cal P}_{2{\bf \lambda}_{1}}^{{\bf \lambda}_{1}{\bf
\lambda}_{1}}\right)^{ab}_{~cd}
{}~-~q^{-1}~\left({\cal P}_{{\bf \lambda}_{2}}^{{\bf \lambda}_{1}
{\bf \lambda}_{1}}\right)^{ab}_{~cd} \eqno (2.6) $$

\noindent
Similarly, in the direct product spaces of ${\bf \lambda}_{N-1}\otimes
{\bf \lambda}_{N-1}$, ${\bf \lambda}_{1}\otimes {\bf \lambda}_{N-1}$,
and ${\bf \lambda}_{N-1}\otimes {\bf \lambda}_{1}$, the solutions
of the simple Yang-Baxter equation are as follows, respectively:
$$\begin{array}{l}
\left(\hat{R}_{q}^{{\bf \lambda}_{N-1}{\bf \lambda}_{N-1}}
\right)^{\bar{a}\bar{b}}_{~\bar{c}\bar{d}}~=~
q~\left({\cal P}_{2{\bf \lambda}_{N-1}}^{{\bf \lambda}_{N-1}{\bf
\lambda}_{N-1}}\right)^{\bar{a}\bar{b}}_{~\bar{c}\bar{d}}
{}~-~q^{-1}~\left({\cal P}_{{\bf \lambda}_{N-2}}^{{\bf \lambda}_{N-1}
{\bf \lambda}_{N-1}}\right)^{\bar{a}\bar{b}}_{~\bar{c}\bar{d}} \\[1mm]
\left(\hat{R}_{q}^{{\bf \lambda}_{1}{\bf \lambda}_{N-1}}
\right)^{\bar{a}b}_{~c\bar{d}}~=~q~\left({\cal P}_{{\bf \lambda}_{1}
+{\bf \lambda}_{N-1}}^{{\bf \lambda}_{1}{\bf \lambda}_{N-1}}
\right)^{\bar{a}b}_{~c\bar{d}}~+~(-q)^{1-N}~\left({\cal P}_{{\bf 0}}^{
{\bf \lambda}_{1}{\bf \lambda}_{N-1}}\right)^{\bar{a}b}_{~c\bar{d}}  \\[1mm]
\left(\hat{R}_{q}^{{\bf \lambda}_{N-1}{\bf \lambda}_{1}}
\right)^{a\bar{b}}_{~\bar{c}d}~=~q~\left({\cal P}_{{\bf \lambda}_{1}+
{\bf \lambda}_{N-1}}^{{\bf \lambda}_{N-1}{\bf \lambda}_{1}}
\right)^{a\bar{b}}_{~\bar{c}d}~+~(-q)^{1-N}~\left({\cal P}_{{\bf
0}}^{{\bf \lambda}_{N-1}{\bf \lambda}_{1}}
\right)^{a\bar{b}}_{~\bar{c}d}
\end{array} \eqno (2.7) $$

The superscripts, for example ${\bf \lambda}_{1}{\bf \lambda}_{1}$,
have been implied in the super- and sub-scripts $ab$ and $cd$, and
can be neglected. Now, through straightforward calculation, we obtain:
$$\begin{array}{l}
\left(\hat{R}_{q}\right)^{ab}_{~cd}~=~
\left(\hat{R}_{q}\right)^{cd}_{~ab}~=~
\left(\hat{R}_{q}\right)^{\bar{b}\bar{a}}_{~\bar{d}\bar{c}}~=~
\left(\hat{R}_{q}\right)^{\bar{d}\bar{c}}_{~\bar{b}\bar{a}} \\
{}~~~~~~~~=~\left\{\begin{array}{ll} q ~~~~~&{\rm when}~~a=b=c=d \\
\lambda &{\rm when}~~a=c<b=d \\
1&{\rm when}~~a=d\neq b=c \\
0&{\rm the ~else~ cases} \end{array} \right.
\end{array} $$
$$\begin{array}{l}
\left(\hat{R}_{q}^{-1}\right)^{ab}_{~cd}~=~
\left(\hat{R}_{q}^{-1}\right)^{cd}_{~ab}~=~
\left(\hat{R}_{q}^{-1}\right)^{\bar{b}\bar{a}}_{~\bar{d}\bar{c}}~=~
\left(\hat{R}_{q}^{-1}\right)^{\bar{d}\bar{c}}_{~\bar{b}\bar{a}}\\
{}~~~~~~~~=~\left\{\begin{array}{ll} q^{-1} ~~~~~&{\rm when}~~a=b=c=d \\
-~\lambda &{\rm when}~~a=c>b=d \\
1&{\rm when}~~a=d\neq b=c \\
0&{\rm the ~else~ cases} \end{array} \right.
\end{array} \eqno (2.8) $$
$$\begin{array}{l}
\left(\hat{R}_{q}\right)^{\bar{a}b}_{~c\bar{d}}~=~
\left(\hat{R}_{q}\right)^{c\bar{d}}_{~\bar{a}b}~=~
\left\{\begin{array}{ll} 1 ~~~~~&{\rm when}~~a=b=c=d \\
-~q\lambda (-q)^{c-b} &{\rm when}~~a=b>c=d \\
q&{\rm when}~~a=d\neq b=c \\
0&{\rm the ~else~ cases} \end{array} \right.
\end{array} $$
$$\begin{array}{l}
\left(\hat{R}_{q}^{-1}\right)^{\bar{a}b}_{~c\bar{d}}~=~
\left(\hat{R}_{q}^{-1}\right)^{c\bar{d}}_{~\bar{a}b}~=~
\left\{\begin{array}{ll} 1 ~~~~~&{\rm when}~~a=b=c=d \\
q^{-1}\lambda (-q)^{c-b} &{\rm when}~~a=b<c=d  \\
q^{-1}&{\rm when}~~a=d\neq b=c \\
0&{\rm the ~else~ cases} \end{array} \right.
\end{array}  $$

\noindent
Hereafter, $\lambda$ and the $q$-number $[m]$ are defined as follows:
$$\begin{array}{c}
\lambda = q-q^{-1},~~~~\lambda^{2}~+~4~=~[2]^{2},~~~~
\lambda^{2}~+~3~=~[3] \\
\lambda^{2}~+~2~=~\displaystyle {[4]\over[2]},
{}~~~~\lambda^{2}~+~1~=~\displaystyle {[6]\over[3][2]} \\
{}~[m]~=~\displaystyle {q^{m}~-~q^{-m} \over q~-~q^{-1}}
\end{array} \eqno (2.9) $$

Remind that in the direct product space of $\lambda_{1}
\otimes \lambda_{N-1}$ there is a subspace belonging to the identity
representation. We define the monoid $\epsilon^{a\bar{b}}$, that is
proportional to the quantum Clebsch-Gordan coefficients ([Ma] P.157)
reducing the product space into the subspace, as follows:
$$\begin{array}{l}
\epsilon^{a\bar{b}}~=~(-1)^{N-1}~\epsilon_{a\bar{b}}~=~
\delta_{ab}~(-1)^{N-a} ~q^{a-(N+1)/2}\\
\epsilon_{\bar{b}a}~=~(-1)^{N-1}~\epsilon^{\bar{b}a}~=~
\delta_{ab}~(-1)^{N-a} ~q^{-a+(N+1)/2}\\
\epsilon^{a\bar{b}}~=~(-1)^{N-1}~D^{a}_{~c}~\epsilon^{\bar{b}c}
{}~=~\epsilon_{\bar{b}c}~D^{c}_{~a}
\end{array} \eqno (2.10) $$
$$\begin{array}{ll}
\epsilon^{a\bar{b}}~\epsilon_{\bar{b}c}~=~\delta^{a}_{c},~~~~
&\epsilon_{\bar{a}b}~\epsilon^{b\bar{c}}~=~\delta^{c}_{a}\\
\epsilon^{a\bar{b}}~\epsilon_{c\bar{b}}~=~(-1)^{N-1}~D^{a}_{~c},~~~~
&\epsilon_{\bar{b}a}~\epsilon^{\bar{b}c}~=~(-1)^{N-1}~(D^{-1})^{c}_{~a}
\end{array} \eqno (2.11)  $$

\noindent
where the diagonal matrix $D$ is related to double antipode action
(see Sec. 3):
$$D^{a}_{~b}~=~\delta^{a}_{b}~q^{-N+2a-1}~=~(-1)^{N-1}~
\epsilon^{a\bar{d}}~\epsilon_{b\bar{d}}
 \eqno (2.12) $$

Four $\hat{R}_{q}$ matrices can be related by $\epsilon^{a\bar{b}}$
matrices:
$$\begin{array}{l}
\left(\hat{R}_{q}^{\pm 1}\right)^{c\bar{d}}_{~\bar{a}b}
{}~=~q^{\pm 1}~\epsilon_{\bar{a}r}~\left(\hat{R}_{q}^{\mp 1}
\right)^{rc}_{~bs}~\epsilon^{s\bar{d}}
{}~=~q^{\pm 1}~\epsilon^{c\bar{r}}~\left(\hat{R}_{q}^{\mp 1}
\right)^{\bar{d}\bar{s}}_{~\bar{r}\bar{a}}~\epsilon_{\bar{s}b} \\
\left(\hat{R}_{q}^{\pm 1}\right)^{\bar{a}b}_{~c\bar{d}}
{}~=~q^{\pm 1}~\epsilon^{\bar{a}r}~\left(\hat{R}_{q}^{\mp 1}
\right)^{bs}_{~rc}~\epsilon_{s\bar{d}}
{}~=~q^{\pm 1}~\epsilon_{c\bar{r}}~\left(\hat{R}_{q}^{\mp 1}
\right)^{\bar{r}\bar{a}}_{~\bar{d}\bar{s}}~\epsilon^{\bar{s}b}
\end{array} \eqno (2.13) $$

$\hat{R}_{q}$ matrix satisfies some important relations [Res]:
$$\left(\hat{R}_{q}^{\pm 1}\right)^{ac}_{~bd}~D^{d}_{~c}~=~
q^{\pm N}~\delta^{a}_{b} \eqno (2.14) $$
$$D^{a}_{~r}~D^{c}_{~s}~\left(\hat{R}_{q}^{\pm 1}\right)^{rs}_{~bd}
{}~=~\left(\hat{R}_{q}^{\pm 1}\right)^{ac}_{~rs}
{}~D^{r}_{~b}~D^{s}_{~d} \eqno (2.15) $$

\vspace{10mm}
\noindent
{\bf 3. Algebra of Functions on the Quantum Group $SU_{q}(N)$}

A quantum group is introduced as the non-commutative and non-cocommutative
Hopf algebra ${\cal A}=Fun_{q}(G)$ obtained by continuous deformations of
the Hopf algebra of the function of a Lie group. The associative algebra
${\cal A}$, the algebra of functions on the quantum group, is freely
generated by non-commutating matrix entries $T^{a}_{~b}$ satisfying the
relation:
$$\begin{array}{c}
\left(\hat{R}_{q}\right)^{ab}_{~rs}~T^{r}_{~c}~T^{s}_{~d}
{}~=~T^{a}_{~r}~T^{b}_{~s}~\left(\hat{R}_{q}\right)^{rs}_{~cd}
\end{array} \eqno (3.1) $$

\noindent
where $\left(\hat{R}_{q}\right)^{ab}_{~cd}$ is given in (2.6).

$T^{a}_{~b}$, the elements of the fundamental representation of
quantum group $SU_{q}(N)$, satisfy the Hopf algebraic relations:
$$\begin{array}{c}
\Delta(T^{a}_{~b})~=~T^{a}_{~d}~\otimes~T^{d}_{~b},~~~
\epsilon(T^{a}_{~b})~=~\delta^{a}_{b} \\
\kappa(T^{a}_{~r})~T^{r}_{~b}~=~T^{a}_{~r}~\kappa(T^{r}_{~b})~
=~\delta^{a}_{b} ,~~~~
\kappa^{2}(T^{a}_{~b})~=~D^{a}_{~c}~T^{c}_{~d}~(D^{-1})^{d}_{~b}
\end{array} \eqno (3.2) $$

\noindent
where the diagonal matrix $D$ is given in (2.12).

The $q$-determinant $det_{q}T$ is commutant with any element $T^{a}_{~b}$.
For quantum group $SU_{q}(N)$ we have:
$$\begin{array}{l}
det_{q}T~=~\displaystyle \sum_{P}~(-q)^{\ell(P)}~T^{1}_{~p_{1}}
{}~\cdots ~T^{N}_{~p_{N}}~=~1,~~~~~\\
q^{*}~=~q,~~~~~
\left(T^{a}_{~b}\right)^{*}~=~\kappa \left(T^{b}_{~a}\right)
\end{array} \eqno (3.3) $$

\noindent
where $\ell(P)$ is the minimal number of inversions in the permutation
$P$. $\kappa \left(T^{b}_{~a}\right)$ can be expressed as antisymmetrized
product of $(N-1)$ $T^{a}_{~b}$ [CSWW]. On the other hand, the conjugate
of the fundamental representation ${\bf \lambda}_{1}$ is equivalent
to the representation ${\bf \lambda}_{N-1}$. Define:
$$T^{\bar{a}}_{~\bar{b}}~=~\epsilon_{\bar{b}r}~\kappa \left(T^{r}_{~s}
\right)~\epsilon^{s\bar{a}},~~~~
\kappa \left(T^{b}_{~a}\right)~=~\epsilon^{b\bar{r}}~T^{\bar{s}}_{~\bar{r}}
{}~\epsilon_{\bar{s}a}  \eqno (3.4) $$

\noindent
For $SU_{q}(2)$, ${\bf \lambda}_{1}={\bf \lambda}_{N-1}$. In this case the
fundamental representation becomes self-conjugate:
$$T^{\bar{a}}_{~\bar{b}}~=~T^{3-a}_{~3-b} $$

{}From (3.1), (3.2) and (2.13) we can show that $T^{\bar{a}}_{~\bar{b}}\in
{\cal A}$ defined in (3.4) belongs to the representation ${\bf
\lambda}_{N-1}$.

\vspace{5mm}
\noindent
{\bf Proposition 1.} {\it $T^{\bar{a}}_{~\bar{b}}\in {\cal A}$
defined in (3.4) satisfy}:
$$\begin{array}{l}
\left(\hat{R}_{q}\right)^{\bar{a}b}_{~r\bar{s}}~T^{r}_{~c}~
T^{\bar{s}}_{~\bar{d}}
{}~=~T^{\bar{a}}_{~\bar{r}}~T^{b}_{~s}~
\left(\hat{R}_{q}\right)^{\bar{r}s}_{~c\bar{d}} \\[1mm]
\left(\hat{R}_{q}\right)^{a\bar{b}}_{~\bar{r}s}~T^{\bar{r}}_{~\bar{c}}~
T^{s}_{~d}~=~T^{a}_{~r}~T^{\bar{b}}_{~\bar{s}}~
\left(\hat{R}_{q}\right)^{r\bar{s}}_{~\bar{c}d} \\[1mm]
\left(\hat{R}_{q}\right)^{~\bar{a}\bar{b}}_{~\bar{r}\bar{s}}~
T^{\bar{r}}_{~\bar{c}}~T^{\bar{s}}_{~\bar{d}}
{}~=~T^{\bar{a}}_{~\bar{r}}~T^{\bar{b}}_{~\bar{s}}~
\left(\hat{R}_{q}\right)^{\bar{r}\bar{s}}_{~\bar{c}\bar{d}}
\end{array} \eqno (3.5) $$

\noindent
{\it Proof.} From (3.1) and (3.2) we have
$$\kappa(T^{a}_{~s})~\left(\hat{R}_{q}\right)^{sb}_{~rd}~T^{r}_{~c}
{}~=~T^{b}_{~s}~\left(\hat{R}_{q}\right)^{as}_{~cr}~\kappa(T^{r}_{~d}) $$

\noindent
In terms of (2.13) and (3.4) we obtain:
$$\begin{array}{rl}
\kappa(T^{a}_{~s})~\left(\hat{R}_{q}\right)^{sb}_{~rd}~T^{r}_{~c}
&=~\epsilon^{a\bar{t}}~T^{\bar{u}}_{~\bar{t}}
{}~\epsilon_{\bar{u}s}~\left(\hat{R}_{q}\right)^{sb}_{~rd}~T^{r}_{~c} \\
&=~q~\epsilon^{a\bar{t}}~\left\{T^{\bar{u}}_{~\bar{t}}
{}~\left(\hat{R}_{q}^{-1}\right)^{b\bar{v}}_{~\bar{u}r}~
T^{r}_{~c}\right\}~\epsilon_{\bar{v}d}~ \\
T^{b}_{~s}~\left(\hat{R}_{q}\right)^{as}_{~cr}~\kappa(T^{r}_{~d})
&=~T^{b}_{~s}~\left(\hat{R}_{q}\right)^{as}_{~cr}~
\epsilon^{r\bar{u}}~T^{\bar{v}}_{~\bar{u}}~\epsilon_{\bar{v}d}\\
&=~q~\epsilon^{a\bar{t}}~\left\{T^{b}_{~s}~\left(\hat{R}_{q}^{-1}
\right)^{s\bar{u}}_{~\bar{t}c}~T^{\bar{v}}_{~\bar{u}}\right\}
{}~\epsilon_{\bar{v}d} \end{array} $$

\noindent
Comparing the quantities in the brackets, we get the first relation
in (3.5). The proof of the other two relations in (3.5) can be
performed analogously. $~~~$ Q.E.D.

The direct product of $T^{a}_{~c}$ and $T^{\bar{b}}_{~\bar{d}}$ spans
the mixed space of the adjoint and identity representations:
$$T^{a}_{~c}~T^{\bar{b}}_{~\bar{d}}
{}~=~T^{a}_{~c}~\epsilon_{\bar{d}r}~\kappa(T^{r}_{~s})~\epsilon^{s\bar{b}}
\eqno (3.6) $$

\noindent
In order to separate the singlet and the adjoint components
in $T^{a}_{~c}T^{\bar{b}}_{~\bar{d}}$, we define the generalized $q$-Pauli
matrices, that obviously have to be proportional to the quantum
Clebsch-Gordan coefficients:
$$\begin{array}{l}
(\sigma^{0})_{a\bar{b}}~=~q^{2}~(\sigma_{0})^{a\bar{b}}
{}~=~-~q~[N]^{-1/2}~\epsilon^{a\bar{b}}\\
(\sigma^{i\bar{j}})_{a\bar{b}}~=~
(\sigma_{i\bar{j}})^{a\bar{b}}~=~(-1)^{N-b}~
\delta_{ai}~\delta_{bj},~~~~i\neq j \\
(\sigma^{j\bar{j}})_{a\bar{b}}~=~\delta_{ab}~\left\{\begin{array}{ll}
(-1)^{N+a}~q^{a-j+(N+1)/2}~\{[j-1][j]\}^{-1/2}~~~&{\rm when}~~a<j\\
(-1)^{N+j+1}~q^{-j+(N+1)/2}~\{[j-1]/[j]\}^{1/2}~~~&{\rm when}~~a=j\\
0 &{\rm when}~~a>j \end{array} \right. ,~~~~j\geq 2\\
(\sigma_{j\bar{j}})^{a\bar{b}}~=~\delta_{ab}~\left\{\begin{array}{ll}
(-1)^{N+a}~q^{a+j-(N+3)/2}~\{[j-1][j]\}^{-1/2}~~~&{\rm when}~~a<j\\
(-1)^{N+j+1}~q^{j-(N+3)/2}~\{[j-1]/[j]\}^{1/2}~~~&{\rm when}~~a=j\\
0 &{\rm when}~~a>j \end{array} \right. ,~~~~j\geq 2
 \end{array} \eqno (3.7) $$

\noindent
where the repeated $j$ is not summed. In the present paper, we describe
the adjoint components by a capital Latin, such as $I$, that runs over
$(i\bar{j})$ where $i\neq j$, and $(j\bar{j})$
where $j\geq 2$, and describe both singlet and adjoint components
by a capital Latin with a hat, like $\hat{I}$, that runs over $0$ and $I$.
In this notation, we have:
$$(\sigma^{\hat{I}})_{a\bar{b}}~(\sigma_{\hat{J}})^{a\bar{b}}~=~
\delta^{\hat{I}}_{\hat{J}},~~~~(\sigma^{\hat{I}})_{a\bar{b}}~
(\sigma_{\hat{I}})^{c\bar{d}}~=~\delta^{c}_{a}~\delta^{d}_{b}
\eqno (3.8) $$

Sometime, another set of $q$-Pauli matrices are useful:
$$\begin{array}{l}
(\sigma^{\hat{I}})_{a}^{~b}~=~(\sigma^{\hat{I}})_{a\bar{d}}
{}~\epsilon^{b\bar{d}},~~~~(\sigma_{\hat{I}})^{a}_{~b}~=~
(\sigma_{\hat{I}})^{a\bar{d}}~\epsilon_{\bar{d}b} \\
(\sigma^{\hat{I}})_{a}^{~b}~(\sigma_{\hat{J}})^{a}_{~b}~=~
\delta^{\hat{I}}_{\hat{J}},~~~~(\sigma^{\hat{I}})_{a}^{~b}~
(\sigma_{\hat{I}})^{c}_{~d}~=~\delta^{c}_{a}~\delta^{b}_{d}
\end{array} \eqno (3.9) $$
$$\begin{array}{l}
(\sigma^{0})_{a}^{~b}~=~-~q~[N]^{-1/2}~D^{b}_{~a},~~~~
(\sigma_{0})^{a}_{~b}
{}~=~-~q^{-1}~[N]^{-1/2}~\delta^{a}_{\bar{b}}\\
(\sigma^{i\bar{j}})_{a}^{~b}~=~\delta_{a}^{i}~\delta_{b}^{j}~
q^{b-(N+1)/2},~~~~
(\sigma_{i\bar{j}})^{a}_{~b}~=~\delta^{a}_{i}~\delta^{b}_{j}~
q^{-b+(N+1)/2},~~~~i\neq j \\
(\sigma^{j\bar{j}})_{a}^{~b}~=~\delta_{ab}~\left\{\begin{array}{ll}
q^{2a-j}~\{[j-1][j]\}^{-1/2}~~~&{\rm when}~~a<j\\
-~\{[j-1]/[j]\}^{1/2}~~~&{\rm when}~~a=j\\
0 &{\rm when}~~a>j \end{array} \right. ,~~~~j\geq 2\\
(\sigma_{j\bar{j}})^{a}_{~b}~=~\delta_{ab}~\left\{\begin{array}{ll}
q^{j-1}~\{[j-1][j]\}^{-1/2}~~~&{\rm when}~~a<j\\
-~q^{-1}~\{[j-1]/[j]\}^{1/2}~~~&{\rm when}~~a=j\\
0 &{\rm when}~~a>j \end{array} \right. ,~~~~j\geq 2
 \end{array} \eqno (3.10) $$

\noindent
where the repeated $j$ is not summed.
When $q \rightarrow 1$, both $(\sigma^{\hat{I}})_{a}^{~b}$ and
$(\sigma_{\hat{I}})^{c}_{~d}$ tend to the usual generalized Pauli
matrices. Now, it can be seen that the singlet and the adjoint components
in $T^{a}_{~c}T^{\bar{b}}_{~\bar{d}}$ are separated explicitly:
$$\begin{array}{l}
M^{\hat{I}}_{~\hat{J}}~=~(\sigma^{\hat{I}})_{a\bar{b}}~
T^{a}_{~c}~T^{\bar{b}}_{~\bar{d}}~(\sigma_{\hat{J}})^{c\bar{d}}
{}~=~(\sigma^{\hat{I}})_{a}^{~d}~T^{a}_{~c}~\kappa(T^{b}_{~d})
{}~(\sigma_{\hat{J}})^{c}_{~b} \\
M^{0}_{~0}~=~{\bf 1 },~~~~M^{I}_{~0}~=~M^{0}_{~I}~=~0 \\
\kappa^{2}(M^{\hat{I}}_{~\hat{J}})~=~D^{\hat{I}}_{~\hat{K}}
{}~M^{\hat{K}}_{~\hat{L}}~(D^{-1})^{\hat{L}}_{~\hat{J}}\\
D_{~0}^{0}~=~1,~~~~~~D_{~(i\bar{j})}^{(i\bar{j})}
{}~=~q^{2(i-j)}
\end{array} \eqno (3.11) $$

\noindent
where $D$ is a diagonal matrix.

The linear functionals $(L^{\pm})^{a}_{~b}$, defined by their
values on the entries $T^{a}_{~b}$, belong to the dual Hopf algebra
${\cal A}'$ [FRT]:
$$(L^{+})^{a}_{~b}\left(T^{c}_{~d}\right)~=~q^{-1/N}~
(\hat{R}_{q})^{ac}_{~db}~,~~~~
(L^{-})^{a}_{~b}\left(T^{c}_{~d}\right)~=~q^{1/N}~
(\hat{R}^{-1}_{q})^{ac}_{~db} \eqno (3.12) $$
$$\begin{array}{c}
\Delta'\left( (L^{\pm})^{a}_{~b} \right)~=~(L^{\pm})^{a}_{~c}
{}~\otimes~(L^{\pm})^{c}_{~b},~~~~
\epsilon'\left( (L^{\pm})^{a}_{~b}\right)~=~\delta^{a}_{b} \\
\kappa'\left( (L^{\pm})^{a}_{~c}\right)~\left(L^{\pm}\right)^{c}_{~b}
{}~=~\delta^{a}_{b}~=~
(L^{\pm})^{a}_{~c}~\kappa'\left((L^{\pm})^{c}_{~b}\right)
\end{array} \eqno (3.13) $$
$$\begin{array}{l}
\left(\hat{R}_{q}\right)^{ab}_{~rs}~\left(L^{\pm}\right)^{s}_{~d}
{}~\left(L^{\pm}\right)^{r}_{~c}~=~
\left(L^{\pm}\right)^{b}_{s}~\left(L^{\pm}\right)^{a}_{~r}~
\left(\hat{R}_{q}\right)^{rs}_{~cd} \\
\left(\hat{R}_{q}\right)^{ab}_{~rs}~\left(L^{+}\right)^{s}_{~d}
{}~\left(L^{-}\right)^{r}_{~c}~=~
\left(L^{+}\right)^{b}_{s}~\left(L^{-}\right)^{a}_{~r}~
\left(\hat{R}_{q}\right)^{rs}_{~cd} \end{array}
 \eqno (3.14) $$

{}From (3.12) we have:
$$\begin{array}{l}
\kappa'\left( (L^{\pm})^{a}_{~b}\right)(T^{c}_{~d})~=~q^{\pm 1/N}~\left(
\hat{R}_{q}^{\mp 1}\right)^{ca}_{~bd} \\
\left( (L^{\pm})^{a}_{~b}\right)(T^{\bar{c}}_{~\bar{d}})~=~
q^{\pm 1/N}~\epsilon_{\bar{d}r}~\left(\hat{R}_{q}^{\mp 1}\right)^{ra}_{~bs}~
\epsilon^{s\bar{c}}
{}~=~q^{\mp(1-1/N)}~\left(\hat{R}_{q}^{\pm 1}\right)^{a\bar{c}}_{~\bar{d}b}
\\
\kappa'\left( (L^{\pm})^{a}_{~b}\right)(T^{\bar{c}}_{~\bar{d}})~=~
q^{\mp 1/N}~\epsilon_{\bar{d}r}~D^{r}_{~u}~\left(\hat{R}_{q}^{\pm 1}
\right)^{au}_{~vb}~
(D^{-1})^{v}_{~s}~\epsilon^{s\bar{c}}
\end{array} \eqno (3.15) $$

\vspace{10mm}
\noindent
{\bf 4. Bicovariant Differential Calculus on Quantum Groups $SU_{q}(N)$}

[CSWW] constructed the bimodule $\Gamma$ for the quantum groups
$SU_{q}(N)$ explicitly. In order to avoid confusion with the spacetime
derivative, following Watamura's notation [Wat] in the theory of
$q$-deformed BRST algebra, we call the first-order differential
operator on ${\cal A}$ the BRST transformation operator, denoted
by $\delta$:
$$\begin{array}{l}
\delta: ~{\cal A}~\rightarrow~\Gamma \\
\rho~=~\sum~\alpha~\delta~ \beta~\in~\Gamma,~~~~{\rm if}~~\rho \in \Gamma
\end{array} \eqno (4.1) $$

\noindent
where $\alpha, ~\beta ~\in ~{\cal A}$.

A left action $\Delta_{L}$ and a right action $\Delta_{R}$ of the
quantum group on $\Gamma$ are defined as follows:
$$\begin{array}{ll}
\Delta_{L}~:~\Gamma~\rightarrow~{\cal A}~\otimes ~\Gamma,~~~~
&\Delta_{L}(\alpha \delta \beta)~=~\Delta(\alpha)~( id~ \otimes ~\delta)
{}~\Delta(\beta) \\
\Delta_{R}~:~\Gamma~\rightarrow~\Gamma ~\otimes ~{\cal A},~~~~
&\Delta_{R}(\alpha \delta \beta)~=~\Delta(\alpha)~( \delta ~\otimes ~id)
{}~\Delta(\beta)
\end{array} \eqno (4.2) $$

\noindent
The tensor product between elements $\rho$, $\rho'\in \Gamma$
is defined to have the properties:
$$\begin{array}{c}
(\rho~\alpha)~\otimes~\rho'~=~\rho~\otimes~(\alpha~\rho')\\
\alpha(\rho~\otimes~\rho')~=~(\alpha~\rho)~\otimes~\rho',~~~~
(\rho~\otimes~\rho')\alpha~=~\rho~\otimes~(\rho'~\alpha)
\end{array} \eqno (4.3) $$

\noindent
The left and right actions on $\Gamma \otimes \Gamma$ are defined by:
$$\begin{array}{l}
\Delta_{L}~:~\Gamma~\otimes~\Gamma~\rightarrow~{\cal A}~\otimes~
\Gamma~\otimes~\Gamma \\
\Delta_{R}~:~\Gamma~\otimes~\Gamma~\rightarrow~\Gamma~
\otimes~\Gamma~\otimes ~{\cal A} \end{array} \eqno (4.4) $$

\noindent
For example,
$$\begin{array}{l}
\Delta_{L}(\rho_{1})~=~\alpha_{1}~\otimes~\rho_{1}',~~~
\Delta_{L}(\rho_{2})~=~\alpha_{2}~\otimes~\rho_{2}' \\
\Delta_{L}(\rho_{1}~\otimes ~\rho_{2})~=~\alpha_{1}~\alpha_{2}~
\otimes~\rho_{1}'~\otimes~\rho_{2}' \end{array}\eqno (4.5) $$

\noindent
{}From the definition we have:
$$\begin{array}{rl}
(\epsilon~ \otimes ~id)~\Delta_{L}(\rho)~=~\rho,~&~
(id~ \otimes ~\epsilon)~\Delta_{R}(\rho)~=~\rho \\
(\Delta ~\otimes ~id)~\Delta_{L}~=~(id ~\otimes~ \Delta_{L})~\Delta_{L},~&~
(id ~\otimes~ \Delta)~\Delta_{R}~=~(\Delta_{R} ~\otimes~ id)~\Delta_{R} \\
(id~ \otimes~ \Delta_{R} )~\Delta_{L}&=~(\Delta_{L}~ \otimes~ id)
{}~\Delta_{R}$$
\end{array} \eqno (4.6) $$

[CSWW] constructed the fundamental bicovariant bimodule of
$SU_{q}(N)$:
$$\begin{array}{l}
\Delta_{R}(\eta^{a})~=~\eta^{a} ~\otimes ~{\bf 1} ,~~~~
\Delta_{L}(\eta^{a})~=~T^{a}_{~b} ~\otimes ~\eta^{b} \\
\eta^{a}~\alpha~=~(\alpha~ *~ f^{a}_{~b})~\eta^{b},~~~~
\alpha~\eta^{a}~=~\eta^{b}~(\alpha~ *~ f^{a}_{~b}\circ \kappa)
\end{array} \eqno (4.7) $$

\noindent
where $(\alpha * f^{a}_{~b})\equiv (f^{a}_{~b}\otimes id) \Delta \alpha$.

Applying the $*$-operation on (4.7), [CSWW] obtained
$$\begin{array}{l}
(\eta^{a})^{*}~\equiv~\bar{\eta}_{a},~~~~
\Delta_{R}(\bar{\eta}_{a})~=~\bar{\eta}_{a} ~\otimes ~{\bf 1} ,~~~~
\Delta_{L}(\bar{\eta}_{a})~=~\kappa(T^{b}_{~a}) ~\otimes ~\bar{\eta}_{b} \\
\bar{\eta}_{a}~\alpha~=~(\alpha~ *~ \bar{f}^{b}_{~a})~\bar{\eta}_{b},~~~~
\alpha~\bar{\eta}_{a}~=~\bar{\eta}_{b}~(\alpha~ *~ \bar{f}^{b}_{~a}\circ
\kappa) \end{array} \eqno (4.8) $$

{}From the consistency with the commutation relation of the
generators and the requirement for $SU_{q}(N)$ that the
$q$-determinant of $T^{a}_{~b}$ commutes with any element in $\Gamma$,
[CSWW] found that there are two independent functionals such that the
following constructions are performed in a completely parallel way.
Following [CSWW] we choose one of them as follows:
$$f^{a}_{~b}~=~(L^{+})^{a}_{~b},
{}~~~~\bar{f}^{a}_{~b}~=~\kappa'^{-1}\left((L^{-})^{a}_{~b}\right)
\eqno (4.9) $$

Transforming the bases $\bar{\eta}_{a}$ by the monoid
$\epsilon^{a\bar{b}}$, we obtain the bases $\eta^{\hat{I}}$ for the mixed
representation of identity and adjoint representations:
$$\begin{array}{l}
\bar{\eta}^{\bar{b}}~=~\bar{\eta}_{a}~\epsilon^{a\bar{b}},~~~~
\bar{\eta}_{a}~=~\bar{\eta}^{\bar{b}}~\epsilon_{\bar{b}a} \\
\eta^{\hat{I}}~=~(\sigma^{\hat{I}})_{a\bar{b}}~\eta^{a}~\bar{\eta}^{\bar{b}}
{}~=~(\sigma^{\hat{I}})_{a}^{~b}~\eta^{a}~\bar{\eta}_{b}
\end{array} \eqno (4.10) $$

\noindent
In the commutative relations of the $q$-deformed BRST algebra
the components of identity and adjoint representations are mixed.
For the bases $\eta^{\hat{I}}$ we have
$$\begin{array}{l}
\Delta_{R}(\eta^{\hat{J}})~=~\eta^{\hat{J}} ~\otimes ~{\bf 1} ,~~~~
\Delta_{L}(\eta^{\hat{J}})~=~M^{\hat{J}}_{~\hat{K}} ~\otimes ~\eta^{\hat{K}} \\
\alpha ~\eta^{\hat{J}}~=~\eta^{\hat{K}}(\alpha~ *~ L^{~\hat{J}}_{\hat{K}})
,~~~\alpha\in {\cal A},~~~L^{~\hat{J}}_{\hat{K}}\in {\cal A}'
\end{array} \eqno (4.11) $$

\noindent
where
$$\begin{array}{rl}
L^{~\hat{J}}_{\hat{I}}&=~(\sigma^{\hat{J}})_{a}^{~d}~
\kappa' \left((L^{+})^{a}_{~b}\right)
{}~(L^{-})^{c}_{~d}~(\sigma_{\hat{I}})^{b}_{~c} \\
&=~(\sigma^{\hat{J}})_{a\bar{r}}~\left\{\epsilon^{d\bar{r}}~
\kappa' \left((L^{+})^{a}_{~b}\right)
{}~(L^{-})^{c}_{~d}~\epsilon_{\bar{s}c}\right\}~(\sigma_{\hat{I}})^{b\bar{s}}\\
L^{~\hat{J}}_{\hat{I}}(ab)&=~L^{~\hat{K}}_{\hat{I}}(a)~
L^{~\hat{J}}_{\hat{K}}(b),~~~~
L^{~\hat{J}}_{\hat{I}}({\bf 1})~=~\delta^{\hat{J}}_{\hat{I}} \\
(\rho ~*~ L^{~\hat{J}}_{\hat{I}})&=~(L^{~\hat{J}}_{\hat{I}}~ \otimes
{}~id)~\Delta_{L}(\rho)
\end{array} \eqno (4.12) $$

\noindent
Note that:
$$\Delta_{L}(\eta^{0})~=~{\bf 1}~ \otimes ~\eta^{0},~~~~
\Delta_{R}(\eta^{0})~=~\eta^{0}~ \otimes ~{\bf 1} \eqno (4.13) $$

The bases of the left-invariant element of $\Gamma$ are easy to be
calculated from $\eta^{\hat{J}}$:
$$\omega^{\hat{J}}~=~\kappa(M^{\hat{J}}_{~\hat{K}})~\eta^{\hat{K}},
{}~~~~\Delta_{L}(\omega^{\hat{J}})
{}~=~{\bf 1} ~\otimes ~\omega^{\hat{J}},~~~~\Delta_{R}(\omega^{\hat{J}})~=~
\omega^{\hat{K}} ~\otimes ~\kappa(M^{\hat{J}}_{~\hat{K}}) \eqno (4.14) $$

As the analogue of the ordinary permutation operator, a bimodule
automorphism $\Lambda$ in $\Gamma \otimes \Gamma$ is defined by:
$$\begin{array}{c}
\Lambda(\omega^{\hat{J}}~\otimes ~\eta^{\hat{K}})~=~
\eta^{\hat{K}}~\otimes ~\omega^{\hat{J}} \\
\Lambda(\alpha~\tau)~=~\alpha~\Lambda(\tau),~~~~
\Lambda(\tau \alpha)~=~\Lambda(\tau)~\alpha, ~~~~ \alpha\in {\cal A},~~\tau\in
\Gamma\otimes \Gamma \end{array}  \eqno (4.15) $$

\noindent
Thus, we have
$$\Lambda(\eta^{\hat{I}}\otimes \eta^{\hat{J}})~=~\Lambda^{\hat{I}
\hat{J}}_{~\hat{K}\hat{L}}
{}~\eta^{\hat{K}}\otimes \eta^{\hat{L}},~~~~\Lambda^{\hat{I}
\hat{J}}_{~\hat{K}\hat{L}}~=~L^{~\hat{J}}_{\hat{K}}
(M^{\hat{I}}_{~\hat{L}}) \eqno (4.16) $$

In terms of (2.13), (3.12) and (3.15) we are able to compute
$\Lambda^{\hat{I}\hat{J}}_{~~\hat{K}\hat{L}}$ explicitly:
$$\begin{array}{l}
\left(\sigma_{\hat{I}}\right)^{a\bar{b}}~
\left(\sigma_{\hat{J}}\right)^{c\bar{d}}~
\Lambda^{\hat{I}\hat{J}}_{~~\hat{K}\hat{L}}~
\left(\sigma^{\hat{K}}\right)_{i\bar{j}}~
\left(\sigma^{\hat{L}}\right)_{k\bar{\ell}}~
{}~=~\left(\sigma^{\hat{K}}\right)_{i\bar{j}}~
L_{\hat{K}}^{~\hat{S}}(T^{a}_{~k})~
L_{\hat{S}}^{~\hat{J}}(T^{\bar{b}}_{~\bar{\ell}})~
\left(\sigma_{\hat{J}}\right)^{c\bar{d}} \\
{}~~~~=\kappa'\left( (L^{+})^{r}_{~i} \right)(T^{a}_{~u})~
\epsilon_{\bar{j}t}~(L^{-})^{t}_{~w} (T^{u}_{~k})~\epsilon^{w\bar{s}}~
\kappa'\left( (L^{+})^{c}_{~r} \right)(T^{\bar{b}}_{~\bar{v}})~
\epsilon_{\bar{s}x}~(L^{-})^{x}_{~y} (T^{\bar{v}}_{~\bar{\ell}})
{}~\epsilon^{y\bar{d}}\\
{}~~~~=~q^{1/N}~\left(\hat{R}_{q}^{-1}\right)^{ar}_{~iu}~
\epsilon_{\bar{j}t}~q^{1/N}~\left(\hat{R}_{q}^{-1}\right)^{tu}_{~kw}~
{}~\epsilon^{w\bar{s}}~q^{-1/N}~\epsilon_{\bar{v}r'}~D^{r'}_{~s'}
{}~\left(\hat{R}_{q}\right)^{cs'}_{~u'r}~ (D^{-1})^{u'}_{~v'}
{}~\epsilon^{v'\bar{b}}\\
{}~~~~~~~\cdot~\epsilon_{\bar{s}x}~q^{1-1/N}~\left(\hat{R}_{q}^{-1}
\right)^{x\bar{v}}_{~\bar{\ell}y}
{}~\epsilon^{y\bar{d}}\\
{}~~~~=~q~\epsilon_{\bar{j}t}~\epsilon_{s'\bar{v}}
{}~\epsilon^{\bar{b}u'}~~\epsilon^{y\bar{d}}
{}~\left(\hat{R}_{q}^{-1}\right)^{ar}_{~iu}~
\left(\hat{R}_{q}^{-1}\right)^{tu}_{~kw}~
\left(\hat{R}_{q}\right)^{cs'}_{~u'r}~
\left(\hat{R}_{q}^{-1}\right)^{w\bar{v}}_{~\bar{\ell}y}\\
{}~~~~=~\left(\hat{R}_{q}^{-1}\right)^{ar}_{~iu}~
\left(\hat{R}_{q}\right)^{u\bar{s}}_{~\bar{j}k}~
\left(\hat{R}_{q}^{-1}\right)^{\bar{b}c}_{~r\bar{v}}~
\left(\hat{R}_{q}\right)^{\bar{v}\bar{d}}_{~\bar{s}\bar{\ell}}
\end{array} \eqno (4.17) $$

For given matrices $({\cal P}_{1})^{ab}_{~ij}$ and
$({\cal P}_{2})^{\bar{c}\bar{d}}_{~\bar{k}\bar{\ell}}$
define a operator $({\cal P}_{1},{\cal P}_{2})^{\hat{I}
\hat{J}}_{~\hat{K}\hat{L}}$ [CSWW]:
$$\begin{array}{rl}
({\cal P}_{1},{\cal P}_{2})^{\hat{I}\hat{J}}_{~\hat{K}\hat{L}}
&=~(\sigma^{\hat{I}})_{a\bar{b}}~(\sigma^{\hat{J}})_{c\bar{d}}~
\left(\hat{R}_{q}^{-1}\right)^{\bar{b}c}_{~r\bar{v}}~
(X)^{ar}_{~iu}~(Y)^{\bar{v}\bar{d}}_{~\bar{s}\bar{\ell}}
\left(\hat{R}_{q}\right)^{u\bar{s}}_{~\bar{j}k}~
(\sigma_{\hat{K}})^{i\bar{j}}~(\sigma_{\hat{L}})^{k\bar{\ell}}
 \end{array} \eqno (4.18) $$

\noindent
Hence, we obtain four projection operators, orthogonal to each other:
$$\left({\cal P}_{2{\bf \lambda}_{1}},
{\cal P}_{2{\bf \lambda}_{N-1}}\right),~~~~
\left({\cal P}_{2{\bf \lambda}_{1}},{\cal P}_{{\bf \lambda}_{N-2}}\right),~~~~
\left({\cal P}_{{\bf \lambda}_{2}},{\cal P}_{2{\bf \lambda}_{N-1}}\right),~~~~
\left({\cal P}_{{\bf \lambda}_{2}},{\cal P}_{{\bf \lambda}_{N-2}}\right)$$
$$\begin{array}{c}
{\cal P}_{S}~\equiv~\left({\cal P}_{2{\bf \lambda}_{1}},
{\cal P}_{2{\bf \lambda}_{N-1}}\right)~+~
\left({\cal P}_{{\bf \lambda}_{2}},{\cal P}_{{\bf \lambda}_{N-2}}\right)\\
{\cal P}_{A}~\equiv~
\left({\cal P}_{2{\bf \lambda}_{1}},{\cal P}_{{\bf \lambda}_{N-2}}\right)~+~
\left({\cal P}_{{\bf \lambda}_{2}},{\cal P}_{2{\bf \lambda}_{N-1}}\right)\\
{\cal P}_{S}~+~{\cal P}_{A}~=~{\bf 1} \end{array} \eqno (4.19) $$

\noindent
{}From (4.17), (2.6) and (2.7) we have
$$\begin{array}{l}
\Lambda~=~\left(\hat{R}^{-1},\hat{R}\right)\\
{}~~~~=~\left({\cal P}_{2{\bf \lambda}_{1}},
{\cal P}_{2{\bf \lambda}_{N-1}}\right)-~q^{-2}~
\left({\cal P}_{2{\bf \lambda}_{1}},{\cal P}_{{\bf \lambda}_{N-2}}\right)
{}~-~q^{2}~\left({\cal P}_{{\bf \lambda}_{2}},
{\cal P}_{2{\bf \lambda}_{N-1}}\right)~+~
\left({\cal P}_{{\bf \lambda}_{2}},{\cal P}_{{\bf \lambda}_{N-2}}\right)\\
\Lambda^{-1}
{}~=~\left(\hat{R},\hat{R}^{-1}\right)\\
{}~~~~=~\left({\cal P}_{2{\bf \lambda}_{1}},
{\cal P}_{2{\bf \lambda}_{N-1}}\right)-~q^{2}~
\left({\cal P}_{2{\bf \lambda}_{1}},{\cal P}_{{\bf \lambda}_{N-2}}\right)
{}~-~q^{-2}~\left({\cal P}_{{\bf \lambda}_{2}},
{\cal P}_{2{\bf \lambda}_{N-1}}\right)~+~
\left({\cal P}_{{\bf \lambda}_{2}},{\cal P}_{{\bf \lambda}_{N-2}}\right)
\end{array} \eqno (4.20) $$

\noindent
Therefore, $\Lambda^{\hat{I}\hat{J}}_{~\hat{K}\hat{L}}$ satisfy the
Yang-Baxter equation:
$$\Lambda^{\hat{I}\hat{J}}_{~\hat{L}\hat{S}}
{}~\Lambda^{\hat{S}\hat{K}}_{~\hat{T}\hat{R}}
{}~\Lambda^{\hat{L}\hat{T}}_{~\hat{P}\hat{Q}}~=~
\Lambda^{\hat{J}\hat{K}}_{~\hat{L}\hat{S}}~
\Lambda^{\hat{I}\hat{L}}_{~\hat{P}\hat{T}}
{}~\Lambda^{\hat{T}\hat{S}}_{~\hat{Q}\hat{R}}
\eqno (4.21) $$

\noindent
{}From (4.20) we know that the eigenvalues of $\Lambda$ matrix are
$1$, $-q^{2}$ and $-q^{-2}$:
$$\left(\Lambda~+~q^{2}\right)~\left(\Lambda~+~q^{-2}\right)
{}~\left(\Lambda~-~{\bf 1}\right)~=~0 \eqno (4.22) $$

{}From the symmetry of the quantum Clebsch-Gordan coefficients
([Ma] P.156) we have:
$$\left(\Lambda^{-1}\right)^{(i\bar{j})(k\bar{\ell})}_{~(a\bar{b})
(c\bar{d})}~=~\Lambda^{(\ell\bar{k})(j\bar{i})}_{~(d\bar{c})
(b\bar{a})} \eqno (4.23) $$

\noindent
Through direct calculation of (4.17), we obtain the non-vanishing
components of $\Lambda^{\hat{I}\hat{J}}_{~\hat{K}\hat{L}}$ as follows:
$$\begin{array}{c}
\Lambda^{IJ}_{~KL}~=~(\Lambda^{-1})^{IJ}_{~KL}
{}~=~\delta^{I}_{K}~\delta^{J}_{L}~+~f^{IJ}_{P}~\bar{f}_{KL}^{P},~~~
\Lambda^{00}_{~~00}~=~(\Lambda^{-1})^{00}_{~~00}~=~1 \\
\Lambda^{I0}_{~JK}~=~(\Lambda^{-1})^{0I}_{~JK}~=~\lambda~f_{JK}^{I},~~~
\Lambda^{JK}_{~0I}~=~(\Lambda^{-1})^{JK}_{~I0}~=~\lambda~\bar{f}^{JK}_{I} \\
\Lambda^{0I}_{~J0}~=~(\Lambda^{-1})^{I0}_{~0J}~=~\delta^{I}_{J} ,~~~
\Lambda^{I0}_{~0J}~=~(\Lambda^{-1})^{0I}_{~J0}~=~
(\lambda^{2}+1)~\delta^{I}_{J}  \end{array} \eqno (4.24) $$

\noindent
where $f_{JK}^{I}$ and $\bar{f}_{I}^{JK}$ satisfy:
$$\begin{array}{l}
f_{(i\bar{j})(k\bar{\ell})}^{(r\bar{s})}~=~0,~~~~
\bar{f}^{(i\bar{j})(k\bar{\ell})}_{(r\bar{s})}~=~0, \\
{}~~~~~~~~{\rm if}~~
\lambda_{i}-\lambda_{i-1}-\lambda_{j}+\lambda_{j-1}
+\lambda_{k}-\lambda_{k-1}-\lambda_{\ell}+\lambda_{\ell-1}
\neq \lambda_{r}-\lambda_{r-1}-\lambda_{s}+\lambda_{s-1}\\
f_{(i\bar{j})(k\bar{\ell})}^{(r\bar{s})}~=~
f_{(\ell \bar{k})(j\bar{i})}^{(s \bar{r})},~~~~
\bar{f}^{(i\bar{j})(k\bar{\ell})}_{(r\bar{s})}~=~
\bar{f}^{(\ell \bar{k})(j\bar{i})}_{(s \bar{r})},~~~~
f^{I}_{RS}~\bar{f}_{J}^{RS}
{}~=~-~(\lambda^{2}+2)~\delta^{I}_{J}
\end{array} \eqno (4.25) $$

\noindent
The non-vanishing components of $f_{JK}^{I}$ and $\bar{f}_{I}^{JK}$
are listed as follows. In the following (4.26) there is no summation
for the repeated indices.
$$\begin{array}{l}
f_{(i\bar{k})(k\bar{j})}^{(i\bar{j})}~=~[N]^{-1/2}~q^{-k-(N-3)/2},~~~~
f_{(k \bar{j})(i\bar{k})}^{(i \bar{j})}~=~-~[N]^{-1/2}~q^{k-(N-1)/2} \\
\bar{f}^{(i\bar{k})(k\bar{j})}_{(i\bar{j})}~=~-~[N]^{-1/2}~q^{-k+(3N-1)/2},~~~~
\bar{f}^{(k \bar{j})(i\bar{k})}_{(i \bar{j})}~=~[N]^{-1/2}~q^{k-(N+3)/2} \\
{\rm when}~~i~\neq ~j~\neq~k~\neq i \end{array} \eqno (4.26a) $$
$$f_{(k\bar{k})(i\bar{j})}^{(i\bar{j})}~=~\left\{\begin{array}{ll}
-~q^{-N}~\left( \displaystyle{[k-1] \over [N][k]}\right)^{1/2}~~~~~
&{\rm if}~~k=i<j \\
-~q^{1-N}~\left( \displaystyle{[k] \over [N][k-1]}\right)^{1/2}~~~~~
&{\rm if}~~j<i=k\\
q^{2k-N-1}~\left( \displaystyle{[k] \over [N][k-1]}\right)^{1/2}~~~~~
&{\rm if}~~i<j=k\\
q^{2k-N}~\left( \displaystyle{[k-1] \over [N][k]}\right)^{1/2}~~~~~
&{\rm if}~~k=j<i \\
q^{k-N}~\left( [N][k][k-1]\right)^{-1/2}~~~~~
&{\rm if}~~i<k<j \\
-~q^{k-N}~\left( [N][k][k-1]\right)^{-1/2}~~~~~
&{\rm if}~~j<k<i \end{array} \right.
 \eqno (4.26b) $$
$$\bar{f}^{(k\bar{k})(i\bar{j})}_{(i\bar{j})}~=~\left\{\begin{array}{ll}
q^{2N-2k}~\left( \displaystyle{[k-1] \over [N][k]}\right)^{1/2}~~~~~
&{\rm if}~~k=i<j \\
q^{2N-2k+1}\left( \displaystyle{[k] \over [N][k-1]}\right)^{1/2}
-\lambda q^{N-k}\left( \displaystyle{[N] \over [k][k-1]}\right)^{1/2}
&{\rm if}~~j<i=k\\
-q^{-1}\left( \displaystyle{[k] \over [N][k-1]}\right)^{1/2}-
\lambda q^{N-k}\left( \displaystyle{[N] \over [k][k-1]}\right)^{1/2}
&{\rm if}~~i<j=k\\
-~\left( \displaystyle{[k-1] \over [N][k]}\right)^{1/2}~~~~~
&{\rm if}~~k=j<i \\
-~q^{2N-k}~\left( [N][k][k-1]\right)^{-1/2}~~~~~
&{\rm if}~~i<k<j \\
q^{-k}~\left( [N][k][k-1]\right)^{-1/2}~~~~~
&{\rm if}~~j<k<i \\
-~\lambda~q^{N-k}~\left( \displaystyle{[N] \over [k][k-1]}\right)^{1/2}
&{\rm if}~~i<j<k \\
&{\rm or}~~j<i<k \end{array}
 \right. \eqno (4.26c) $$
$$f_{(i\bar{j})(j\bar{i})}^{(k\bar{k})}~=~\left\{\begin{array}{ll}
-~q^{2-k-j}~\left( \displaystyle{[k-1] \over [N][k]}\right)^{1/2}~~~~~
&{\rm if}~~k=i<j \\
-~q^{3-k-j}~\left( \displaystyle{[k] \over [N][k-1]}\right)^{1/2}~~~~~
&{\rm if}~~j<i=k\\
q^{i-k+1}~\left( \displaystyle{[k] \over [N][k-1]}\right)^{1/2}~~~~~
&{\rm if}~~i<j=k\\
q^{i-k+2}~\left( \displaystyle{[k-1] \over [N][k]}\right)^{1/2}~~~~~
&{\rm if}~~k=j<i \\
q^{i-j-k+2}~\left( [N][k][k-1]\right)^{-1/2}~~~~~
&{\rm if}~~i<k<j \\
-~q^{i-j-k+2}~\left( [N][k][k-1]\right)^{-1/2}~~~~~
&{\rm if}~~j<k<i \end{array} \right. \eqno (4.26d) $$
$$\bar{f}^{(i\bar{j})(j\bar{i})}_{(k\bar{k})}~=~\left\{\begin{array}{ll}
q^{N+k-j-2}~\left( \displaystyle{[k-1] \over [N][k]}\right)^{1/2}~~~~~
&{\rm if}~~k=i<j \\
q^{N+k-j-1} \left( \displaystyle{[k] \over [N][k-1]}\right)^{1/2}
-\lambda q^{2k-j-2} \left( \displaystyle{[N] \over [k][k-1]}\right)^{1/2}
&{\rm if}~~j<i=k\\
-q^{i+k-N-3}\left( \displaystyle{[k] \over [N][k-1]}\right)^{1/2}-
\lambda q^{i-2}\left( \displaystyle{[N] \over [k][k-1]}\right)^{1/2}
&{\rm if}~~i<j=k\\
-~q^{i+k-N-2}~\left( \displaystyle{[k-1] \over [N][k]}\right)^{1/2}~~~~~
&{\rm if}~~k=j<i \\
-~q^{i-j+k+N-2}~\left( [N][k][k-1]\right)^{-1/2}~~~~~
&{\rm if}~~i<k<j \\
q^{i-j+k-N-2}~\left( [N][k][k-1]\right)^{-1/2}~~~~~
&{\rm if}~~j<k<i \\
-~\lambda~q^{i-j+k-2}~\left( \displaystyle{[N] \over [k][k-1]}\right)^{1/2}
&{\rm if}~~i<j<k\\
&{\rm or}~~j<i<k
\end{array} \right. \eqno (4.26e) $$
$$\begin{array}{l}
f_{(k\bar{k})(k\bar{k})}^{(k\bar{k})}~=~
\lambda~q^{k-N}~\left( \displaystyle{[k][k-1] \over [N]}\right)^{1/2}\\
\bar{f}^{(k\bar{k})(k\bar{k})}_{(k\bar{k})}~=~
-~\lambda~q^{N-k}~\left\{\left( \displaystyle{[N] \over [k][k-1]}
\right)^{1/2}~-~[N-k]~\left( \displaystyle{[k-1] \over [N][k]}\right)^{1/2}
\right\}\\
\bar{f}^{(j\bar{j})(j\bar{j})}_{(k\bar{k})}~=~
-~\lambda~q^{N-2j+k}~\left( \displaystyle{[N] \over [k][k-1]}\right)^{1/2},
{}~~~~{\rm if}~~j<k\\
\bar{f}^{(j\bar{j})(k\bar{k})}_{(j\bar{j})}~=~
\bar{f}^{(k\bar{k})(j\bar{j})}_{(j\bar{j})}~=~
-~\lambda~q^{N-k}~\left( \displaystyle{[N] \over [k][k-1]}\right)^{1/2},
{}~~~~{\rm if}~~j<k
\end{array} \eqno (4.26f) $$

\noindent
Substituting (4.24) into the Yang-Baxter equation (4.21) for two cases:
$\hat{K}=\hat{P}=$$\hat{Q}=0$, and $\hat{J}=\hat{K}=\hat{P}=0$,
we obtain:
$$\begin{array}{l}
\bar{f}_{T}^{IR}~\bar{f}_{K}^{TS}~f_{RS}^{J}~=~-~(\lambda^{2}+1)~
\bar{f}_{K}^{IJ}\\
f_{RS}^{K}~f_{TJ}^{S}~\bar{f}^{RT}_{I}~=~-~(\lambda^{2}+1)~
f_{IJ}^{K} \end{array} \eqno (4.27) $$

Now, defining the exterior product of the elements in $\Gamma$:
$$\rho~\wedge \rho'~\equiv~\rho~\otimes \rho'~-~\Lambda\left(
\rho~\otimes \rho'\right) \eqno (4.28) $$

\noindent
we have:
$$\eta^{\hat{I}}~\wedge~\eta^{\hat{J}}~=~\left(\delta^{\hat{I}}_{\hat{K}}
{}~\delta^{\hat{J}}_{\hat{L}}~-~\Lambda^{\hat{I}\hat{J}}_{~\hat{K}\hat{L}}
\right)~\left(\eta^{\hat{K}}~\otimes~\eta^{\hat{L}}\right) $$

\noindent
{}From (4.20) and (4.22) we know that $\eta^{I}~\wedge~\eta^{J}$
is annihilated by the projection operator ${\cal P}_{S}$:
$$\begin{array}{c}
({\cal P}_{S})^{\hat{I}\hat{J}}_{~\hat{K}\hat{L}}~\left(\eta^{\hat{K}}
{}~\wedge~\eta^{\hat{L}}\right)~=~0 \\
\begin{array}{rl}
{\cal P}_{S}&=~[2]^{-2}~\left(\Lambda~+~q^{2}\right)~
\left(\Lambda~+~q^{-2}\right) \\
&=~[2]^{-2}~\left\{\Lambda~+~\Lambda^{-1}~+~(\lambda^{2}+2)~{\bf 1}
\right\} \end{array} \end{array} \eqno (4.29) $$

\noindent
namely,
$$\begin{array}{l}
\eta^{0}~\wedge ~\eta^{0}~=~0 \\
\eta^{0}~\wedge ~\eta^{I}~+~
\eta^{I}~\wedge ~\eta^{0}~=~-~\displaystyle {\lambda \over
\lambda^{2}+2}~f^{I}_{JK}~\eta^{J}~\wedge ~\eta^{K} \\
\bar{f}^{IJ}_{P}~f^{P}_{RS}~\eta^{R}~\wedge ~\eta^{S}~=~
-~(\lambda^{2}+2)~\eta^{I}~\wedge ~\eta^{J}
 \end{array} \eqno (4.30) $$

The projection operator ${\cal P}_{A}$ now can be expressed
as follows:
$${\cal P}_{A}~=~[2]^{-2}~\left\{{\bf 2} ~-~\Lambda~-~\Lambda^{-1}\right\}
\eqno (4.31) $$

It is interesting to notice that we can introduce a projection
operator ${\cal P}_{Adj}$ with only the adjoint components
that projects the product space of two adjoint representations
into the adjoint representation space:
$$\begin{array}{l}
\left({\cal P}_{Adj}\right)^{IJ}_{~KL}~=~\displaystyle {[2]^{2}
\over 2(\lambda^{2}+2)}~\left({\cal P}_{A}\right)^{IJ}_{~KL}
{}~=~-~(\lambda^{2}+2)^{-1}~\bar{f}_{T}^{IJ}~f^{T}_{KL}\\
\left({\cal P}_{Adj}\right)^{IJ}_{~RS}~
\left({\cal P}_{Adj}\right)^{RS}_{~KL}
{}~=~\left({\cal P}_{Adj}\right)^{IJ}_{~KL} \\
\left({\cal P}_{Adj}\right)^{IJ}_{~RS}~
\bar{f}_{K}^{RS}~=~\bar{f}_{K}^{IJ},~~~~
\left({\cal P}_{Adj}\right)^{RS}_{~IJ}~
{}~f^{K}_{RS}~=~f^{K}_{IJ} \\
\left({\cal P}_{Adj}\right)^{IJ}_{~KL}~\left(\eta^{K}~
\wedge~ \eta^{L} \right)~=~\eta^{I} ~\wedge ~\eta^{J}
\end{array} \eqno (4.32) $$

Now, we are in the position to define the BRST transformation $\delta$
on ${\cal A}$ and $\Gamma^{\otimes m}$. $\delta$ is a nilpotent operator:
$$\begin{array}{l}
\delta:~{\cal A}~\rightarrow~\Gamma,~~~~~
\delta: ~\Gamma^{\wedge n}~\rightarrow~\Gamma^{\wedge (n+1)} \\
\delta \alpha~=~\{ ig / \lambda \} ~(\eta^{0} ~\alpha
{}~- ~\alpha ~\eta^{0}),~~~~~~\alpha \in {\cal A}\\
\delta \rho~=~\{ ig / \lambda \} ~\left\{\eta^{0}
{}~\wedge~\rho ~-~(-1)^{n} ~\rho ~\wedge~ ~\eta^{0} \right\},
{}~~~\rho\in \Gamma^{\wedge n} \end{array} \eqno (4.33)  $$

Introduce a functional $\chi \in {\cal A}'$:
$$\begin{array}{l}
\delta \alpha~=~\eta^{\hat{J}} ~(\alpha~ * ~\chi_{\hat{J}}) \\
\chi_{\hat{J}}~=~\displaystyle { ig \over \lambda}~\left(~ \epsilon~
\delta^{0}_{\hat{J}}~-~L^{~0}_{\hat{J}}~\right),
{}~~~~\chi_{\hat{J}}({\bf 1})~=~0  \\
\chi_{\hat{J}}(\alpha \beta)~=~\chi_{\hat{J}}(\alpha)~\epsilon(\beta)~+~
L^{~\hat{K}}_{\hat{J}}(a)~\chi_{\hat{K}}(b)\\
\chi_{\hat{J}}~\chi_{\hat{K}}~=~(\chi_{\hat{J}}~\otimes~\chi_{\hat{K}})~\Delta
\end{array} \eqno (4.34) $$

\noindent
where $\chi_{\hat{J}}$ are the $q$-analogues of the
tangent vectors at the identity element of the group, and
$(\cdot*\chi_{\hat{J}})$ are the analogues of right invariant
vector fields [AC]. $\chi_{\hat{I}}(T^{a}_{~b})$ are proportional
to the $q$-deformed Pauli matrices:
$$\begin{array}{rl}
\chi_{I}(T^{a}_{~b})&=~-~ig~[N]^{-1/2}~q^{1-N+2/N}
{}~\left(\sigma_{I}\right)^{a}_{~b} \\
\chi_{0}(T^{a}_{~b})&=~-~ig~[N]^{-1/2}~q^{1-N+2/N}
{}~\left(\sigma_{0}\right)^{a}_{~b}
{}~+~\{ig/\lambda\}~\left(1-q^{2/N}\right)~\delta^{a}_{b} \\
&=~ig~\left\{q^{-N+2/N}~[N]^{-1}
{}~+~\lambda^{-1}~\left(1-q^{2/N}\right)\right\}~\delta^{a}_{b}
\end{array} \eqno (4.35) $$

The $q$-deformed structure constants can be computed from
(4.34) and (4.24):
$$\begin{array}{rl}
{\bf C}_{\hat{J}\hat{K}}^{~\hat{I}}~\equiv~
\chi_{\hat{J}}(M^{\hat{I}}_{~\hat{K}}),~~~~~
&{\bf C}_{\hat{J}\hat{K}}^{~0}~=~{\bf C}_{\hat{J}0}^{~\hat{K}}~=~0\\
{\bf C}_{0K}^{~J}~=~-~ig\lambda~\delta_{K}^{J},~~~~
&{\bf C}_{JK}^{~I}~=~-~ig~f_{JK}^{I} \end{array} \eqno (4.36) $$

\noindent
Since ${\bf C}_{JK}^{~I}$ are proportional to $f_{JK}^{I}$,
they satisfy the weight conservation condition:
$${\bf C}_{(i\bar{j})(k\bar{\ell})}^{~(r\bar{s})}~=~0,~~~~{\rm if}~~
\lambda_{i}-\lambda_{i-1}-\lambda_{j}+\lambda_{j-1}
+\lambda_{k}-\lambda_{k-1}-\lambda_{\ell}+\lambda_{\ell-1}
\neq \lambda_{r}-\lambda_{r-1}-\lambda_{s}+\lambda_{s-1} \eqno (4.37) $$

The $q$-deformed Cartan-Maurer equation can be derived from (4.33):
$$\begin{array}{rl}
\delta \eta^{0}&=~\{ig/ \lambda\}~\left\{\eta^{0}~
\wedge~\eta^{0}~+~\eta^{0}~\wedge~\eta^{0}\right\} ~=~0\\
\delta \eta^{I}&=~\{ig/ \lambda\}~\left\{\eta^{0}~
\wedge~\eta^{I}~+~\eta^{I}~\wedge~\eta^{0}\right\} \\
&=~\{ig/\lambda\}~\left\{\eta^{0}\otimes \eta^{I}
{}~+~\eta^{I}\otimes \eta^{0}~-~\left(\Lambda^{0I}_{~\hat{J}\hat{K}}~+~
\Lambda^{I0}_{~\hat{J}\hat{K}}\right)~\eta^{\hat{J}}\otimes
\eta^{\hat{K}} \right\}\\
&=~-~ig\lambda~\eta^{0}\otimes \eta^{I}
{}~-~ig~f_{JK}^{I}~\eta^{J}\otimes \eta^{K}\\
&=~{\bf C}_{\hat{J}\hat{K}}^{~I}~\eta^{\hat{J}}~\otimes~\eta^{\hat{K}} \\
&=~\eta^{\hat{J}}~\otimes~\left(\eta^{I}~*~\chi_{\hat{J}}\right)
\end{array} $$

\noindent
namely,
$$\begin{array}{rl}
\delta \eta^{0}&=~{\bf C}_{\hat{J}\hat{K}}^{~0}~\eta^{\hat{J}}
{}~\otimes~\eta^{\hat{K}} ~=~0\\
\delta \eta^{I}&=~{\bf C}_{\hat{J}\hat{K}}^{~I}~\eta^{\hat{J}}
{}~\otimes~\eta^{\hat{K}} \\
&=~\{ig/ \lambda\}~\left\{\eta^{0}~
\wedge~\eta^{I}~+~\eta^{I}~\wedge~\eta^{0}\right\} \\
&=~(\lambda^{2}+2)^{-1}~{\bf C}_{JK}^{~I}~
\left(\eta^{J}\wedge \eta^{K}\right) \end{array}
\eqno (4.38) $$

{}From the condition $\delta^{2} \alpha=0$, the functionals $\chi_{J}$
span the "$q$-deformed Lie algebra":
$$\chi_{\hat{I}}~\chi_{\hat{J}}~-~\Lambda^{\hat{K}\hat{L}}_{~\hat{I}\hat{J}}
{}~\chi_{\hat{K}}~\chi_{\hat{L}}~=~
{\bf C}_{\hat{I}\hat{J}}^{~\hat{K}}~\chi_{\hat{K}} \eqno (4.39) $$

\noindent
Acting (4.39) on $M^{\hat{P}}_{~\hat{S}}$, we obtain the $q$-deformed
Jacobi identities satisfied by the $q$-deformed structure constants:
$${\bf C}_{\hat{I}\hat{R}}^{~\hat{P}}{\bf C}_{\hat{J}\hat{S}}^{~\hat{R}}
{}~-~\Lambda^{\hat{K}\hat{L}}_{~\hat{I}\hat{J}}~
{\bf C}_{\hat{K}\hat{R}}^{~\hat{P}}{\bf C}_{\hat{L}\hat{S}}^{~\hat{R}}
{}~=~{\bf C}_{\hat{I}\hat{J}}^{~\hat{R}}{\bf C}_{\hat{R}\hat{S}}^{~\hat{P}}
\eqno (4.40) $$

\noindent
For the adjoint components we obtain from (4.40):
$$\left({\cal P}_{Adj}\right)^{KL}_{~IJ}~
{\bf C}_{KR}^{~P}{\bf C}_{LS}^{~R}
{}~=~\displaystyle {\lambda^{2}+1 \over \lambda^{2}+2 }~{\bf C}_{IJ}^{~R}
{\bf C}_{RS}^{~P}\eqno (4.41) $$

\noindent
In fact, (4.41) is the same as the second relation in (4.27). Similarly,
acting (4.39) on $T^{a}_{~b}$ we obtain following relations:
$$\chi_{\hat{I}}\left(T_{~d}^{a}\right)~
\chi_{\hat{J}}\left(T_{~b}^{d}\right)
{}~-~\Lambda^{\hat{K}\hat{L}}_{~\hat{I}\hat{J}}~
\chi_{\hat{K}}\left(T_{~d}^{a}\right)~
\chi_{\hat{L}}\left(T_{~b}^{d}\right)
{}~=~{\bf C}_{\hat{I}\hat{J}}^{~\hat{R}}
{}~\chi_{\hat{R}}\left(T_{~b}^{a}\right)
\eqno (4.42) $$
$$\left({\cal P}_{Adj}\right)^{KL}_{~IJ}~
\chi_{K}\left(T^{a}_{~d}\right)~\chi_{L}\left(T^{d}_{~b}\right)
{}~=~\xi ~(\lambda^{2}+2)^{-1}~
{}~{\bf C}_{IJ}^{~R}~\chi_{R}\left(T^{a}_{~b}\right)\eqno (4.43) $$

\noindent
where and hereafter, $\xi$ denotes a constant:
$$\xi~=~q^{2/N}~\left\{1~-~\lambda~q^{-N}~[N]^{-1} \right\} \eqno (4.44) $$

\vspace{10mm}
\noindent
{\bf 5. $q$-deformed BRST Algebra}

Watamura [Wat] investigated the $q$-deformed BRST algebra ${\cal B}$
for $SU_{q}(2)$. The investigation can be generalized into the quantum
groups $SU_{q}(N)$ straightforwardly. We sketch the main results
in our notation.

$\eta^{\hat{I}}$ in the bimodule $\Gamma$ is defined as the ghost field
in the BRST algebra, that has the ghost number 1, but the degree of form
0. The gauge potential $A^{\hat{I}}$ has the degree of form 1, but the
ghost number 0. There are two nilpotent operators in the BRST algebra:
The operator $\delta$ increases the ghost number by one, and the
operator $d$ increases the degree of form by one. Neglecting the
matter field, that is irrelevant to our following discussion,
we are only interested in four fields in the BRST algebra
${\cal B}$: $\eta$, $d\eta$, $A$, and $dA$, which
satisfy the following algebraic relations.

Firstly, we introduce an index $n$ that is equal to the difference between
the degree of form and the ghost number. The indices $n$ for $\eta$,
$d\eta$, $A$ and $dA$ are -1, 0, 1, and 2, respectively.
Both nilpotent operators $\delta$ and $d$ satisfy the Leibniz rule in the
graded sense for the index $n$:
$$\begin{array}{l}
\delta^{2}~=~0,~~~~d^{2}~=~0,~~~~d~\delta~+~\delta~d~=~0 \\
d(XY)~=~(dX)Y~+~(-1)^{n_{x}}~X(dY) \\
\delta(XY)~=~(\delta X)Y~+~(-1)^{n_{x}}~X(\delta Y)
\end{array} \eqno (5.1) $$

\noindent
where $X$, $Y\in {\cal B}$, and
$n_{x}$ is the index of $X$. Both $d$ and $\delta$ are covariant
for the left and right actions: For any element
$X \in {\cal B}$ they satisfy:
$$\begin{array}{l}
\Delta_{L}(\delta X)~=~(id~\otimes ~\delta)\Delta_{L}(X),~~~
\Delta_{L}(dX)~=~(id~\otimes ~d)\Delta_{L}(X) \\
\Delta_{R}(\delta X)~=~(\delta~\otimes ~id)\Delta_{R}(X),~~~
\Delta_{R}(dX)~=~(d~\otimes ~id)\Delta_{R}(X) \end{array}
\eqno (5.2) $$

Secondly, the gauge potentials $A^{\hat{I}}$ are assumed [Wat] to
have similar properties like $\eta^{\hat{I}}$. Hereafter, we neglect
the wedge sign $\wedge$ for simplicity.
$$\begin{array}{l}
({\cal P}_{S})^{\hat{I}\hat{J}}_{~\hat{K}\hat{L}}
{}~\left(A^{\hat{K}}~A^{\hat{L}}\right)~=~0 \\
A^{0}~A^{0}~=~0,~~~~
\left({\cal P}_{Adj}\right)^{IJ}_{~KL}~\left(A^{K}~
A^{L} \right)~=~A^{I} ~A^{J} \\
\{ig/ \lambda \}~\left(A^{0}~A^{I}
{}~+ ~A^{I} ~A^{0} \right)
{}~=~(\lambda^{2}+2)^{-1}~{\bf C}_{JK}^{~I}~A^{J}~A^{K}
\end{array} \eqno (5.3) $$

{}From the consistent conditions [Wat], $d\eta^{\hat{J}}$ and
$dA^{\hat{J}}$ have to satisfy another relation:
$$({\cal P}_{A})^{\hat{I}\hat{J}}_{~\hat{K}\hat{L}}
{}~\left(d\eta^{\hat{K}}~d \eta^{\hat{L}}\right)~=~0,~~~~
({\cal P}_{A})^{\hat{I}\hat{J}}_{~\hat{K}\hat{L}}
{}~\left(dA^{\hat{K}}~dA^{\hat{L}}\right)~=~0 \eqno (5.4)$$

\noindent
namely,
$$\begin{array}{l}
d\eta^{I}~d\eta^{0}~=~d\eta^{0}~d\eta^{I}
{}~=~-~\lambda^{-1}~f^{I}_{JK}~d\eta^{J}~d\eta^{K} \\
dA^{I}~dA^{0}~=~dA^{0}~dA^{I}~=~-~\lambda^{-1}~f^{I}_{JK}~dA^{J}~dA^{K}
\end{array} $$

Thirdly, the gauge potential is introduced in the covariant derivative.
The covariant condition of the covariant derivative in the BRST
transformation requires:
$$\delta A^{0}~=~0,~~~~
\delta A^{I}~=~d \eta^{I}~+~\displaystyle { ig \over \lambda}~
\left( \eta^{0}~A^{I}~+~A^{I}~\eta^{0} \right)
 \eqno (5.5) $$

Fourthly, for two different fields $X^{\hat{J}}$ and $Y^{\hat{K}}$
in ${\cal B}$ with indices $n_{x}$ and $n_{y}$, $n_{x}>n_{y}$,
respectively, the consistent condition requires the
following commutative relations:
$$(-1)^{n_{x}n_{y}}~X^{\hat{I}}~Y^{\hat{J}}~=~
Y^{\hat{K}}~\left(X^{\hat{I}}~*~L^{\hat{J}}_{\hat{K}}\right)
{}~=~\Lambda^{\hat{I}\hat{J}}_{~\hat{K}\hat{L}}~Y^{\hat{K}}~X^{\hat{L}}
\eqno (5.6)$$

\noindent
{}From (4.24) we have
$$\begin{array}{l}
(-1)^{n_{x}n_{y}}~X^{0}~Y^{\hat{J}}~=~Y^{\hat{J}}~X^{0} \\
\{ig/\lambda\}~\left( Y^{0}X^{I}~-~(-1)^{n_{x}n_{y}}~X^{I}Y^{0}\right)\\
{}~~~~~~~=~-~ig\lambda~Y^{0}~X^{I}~-~ig~f_{JK}^{~I}~Y^{J}~X^{K}
{}~=~Y^{\hat{J}}X^{\hat{K}}~{\bf C}_{\hat{J}\hat{K}}^{~I}
\end{array} \eqno (5.7) $$

At last, the gauge fields $F^{J}$ satisfy:
$$\begin{array}{c}
F^{\hat{J}}~=~dA^{\hat{J}}~+~\{ig/ \lambda\}~\left(
A^{0}~A^{\hat{J}}~+~A^{\hat{J}}~A^{0}\right)\\
F^{0}~=~dA^{0},~~~~F^{I}~=~dA^{I}~+~(\lambda^{2}+2)^{-1}
{}~{\bf C}_{JK}^{~I}~A^{J}~A^{K} \\
F^{\hat{I}}~\eta^{\hat{J}}~=~\eta^{\hat{K}}~F^{\hat{L}}
{}~\Lambda^{\hat{I}\hat{J}}_{~\hat{K}\hat{L}}
 \end{array} \eqno (5.8) $$
$$\begin{array}{rl}
\delta F^{\hat{I}}&=~\{ig/ \lambda \}
{}~\left( \eta^{0}~F^{\hat{I}}~-~F^{\hat{I}}~\eta^{0}\right)
{}~=~\eta^{\hat{J}}~F^{\hat{K}}~{\bf C}_{\hat{J}\hat{K}}^{~\hat{I}}\\
\end{array} \eqno (5.9) $$
$$\begin{array}{rl}
d F^{\hat{I}}&=~-~\{ig/ \lambda \}
{}~\left( A^{0}~F^{\hat{I}}~-~F^{\hat{I}}~A^{0}\right) \\
&=~-~\{ig/ \lambda \}
{}~\left( A^{0}~dA^{\hat{I}}~-~dA^{\hat{I}}~A^{0}\right) \\
&=~-~A^{\hat{J}}~dA^{\hat{K}}~{\bf C}_{\hat{J}\hat{K}}^{~\hat{I}}
 \end{array} \eqno (5.10) $$

The commutative relation (5.6) can be rewritten as follows.

\vspace{5mm}
\noindent
{\bf Proposition 2.} {\it The components $X^{a}_{~b}$ and $Y^{a}_{~b}$
of two different fields $X^{\hat{J}}$ and $Y^{\hat{K}}$ in ${\cal B}$
with indices $n_{x}$ and $n_{y}$, $n_{x}>n_{y}$, respectively,
satisfy:}

i)
$$(-1)^{n_{x}n_{y}}~X^{a}_{~r}\left(\hat{R}_{q}^{-1}\right)^{ri}_{~sk}
{}~Y^{s}_{~t} ~\left(\hat{R}_{q}^{-1}\right)^{tk}_{~b j}
{}~=~\left(\hat{R}_{q}^{-1}\right)^{ai}_{~rk}~Y^{r}_{~s}
\left(\hat{R}_{q}^{-1}\right)^{sk}_{~tj}~X^{t}_{~b}
\eqno (5.11) $$

ii)
$$\begin{array}{l}
\{ig/\lambda\}~\left( Y^{0}X^{I}~-~(-1)^{n_{x}n_{y}}~X^{I}Y^{0}\right)~
\left(\sigma_{I}\right)^{a}_{~b}\\
{}~~~~~~=~-~ig~q^{1-N}~[N]^{-1/2}~\left( Y^{a}_{~d}~X^{d}_{~b}~-~
(-1)^{n_{x}n_{y}}~X^{a}_{~d}~Y^{d}_{~b}\right)
\end{array} \eqno (5.12) $$

\noindent
{\it where}
$$X^{a}_{~b}~=~X^{\hat{I}}~\left(\sigma_{\hat{I}}\right)^{a}_{~b},~~~~
Y^{a}_{~b}~=~Y^{\hat{I}}~\left(\sigma_{\hat{I}}\right)^{a}_{~b} $$

\noindent
{\it Proof}. (5.6) can be rewritten in terms of the explicit form (4.17) of
$\Lambda^{\hat{I}\hat{J}}_{~\hat{K}\hat{L}}$:
$$\begin{array}{l}
(-1)^{n_{x}n_{y}}~\left(\sigma_{\hat{I}}\right)^{a\bar{b}}~
\left(\sigma_{\hat{J}}\right)^{c\bar{d}}~X^{\hat{I}}~Y^{\hat{J}}\\
{}~~~~~~=~\left(\hat{R}_{q}^{-1}\right)^{ar}_{~iu}~
\left(\hat{R}_{q}\right)^{u\bar{s}}_{~\bar{j}k}~
\left(\hat{R}_{q}^{-1}\right)^{\bar{b}c}_{~r\bar{v}}~
\left(\hat{R}_{q}\right)^{\bar{v}\bar{d}}_{~\bar{s}\bar{\ell}}~
\left(\sigma_{\hat{K}}\right)^{i\bar{j}}~\left(\sigma_{\hat{L}}
\right)^{k\bar{\ell}}~Y^{\hat{K}}~X^{\hat{L}}
\end{array} \eqno (5.13) $$

\noindent
Moving two factors $\left(\hat{R}_{q}^{-1}\right)^{\bar{b}c}_{~r\bar{v}}~
\left(\hat{R}_{q}\right)^{\bar{v}\bar{d}}_{~\bar{s}\bar{\ell}}$ from
the right hand side of (5.13) to the left, and left multiplying (5.13) by
$q^{-1}\epsilon_{\bar{s}s'}\epsilon_{\bar{\ell}\ell'}$, we obtain
the left hand side of the equation as follows:
$$\begin{array}{l}
(-1)^{n_{x}n_{y}}~q^{-1}~\epsilon_{\bar{s}s'}~\epsilon_{\bar{\ell}\ell'}
{}~\left(\hat{R}_{q}^{-1}\right)^{\bar{s}\bar{\ell}}_{~\bar{v}\bar{d}}~
\left(\hat{R}_{q}\right)^{r\bar{v}}_{~\bar{b}c}~
\left(\sigma_{\hat{I}}\right)^{a\bar{b}}~
\left(\sigma_{\hat{K}}\right)^{c\bar{d}}~
X^{\hat{I}}~Y^{\hat{J}}\\
{}~~~~=~(-1)^{n_{x}n_{y}}~q^{-1}~\epsilon_{\bar{s}s'}
{}~\left(\hat{R}_{q}^{-1}\right)^{br}_{~cv}~
\left(\hat{R}_{q}\right)^{v\bar{s}}_{~\bar{d}\ell'}~
X^{a}_{~b}~Y^{c}_{~d'}~\epsilon^{d'\bar{d}}\\
{}~~~~=~(-1)^{n_{x}n_{y}}~X^{a}_{~b}\left(\hat{R}_{q}^{-1}\right)^{br}_{~cv}
{}~Y^{c}_{~d} ~\left(\hat{R}_{q}^{-1}\right)^{dv}_{~\ell' s'}~
\end{array} $$

\noindent
The right hand side becomes:
$$\begin{array}{l}
q^{-1}~\epsilon_{\bar{s}s'}~\epsilon_{\bar{\ell}\ell'}
{}~\left(\hat{R}_{q}^{-1}\right)^{ar}_{~iu}~
\left(\hat{R}_{q}\right)^{u\bar{s}}_{~\bar{j}k}~
\left(\sigma_{\hat{K}}\right)^{i\bar{j}}~
\left(\sigma_{\hat{L}}\right)^{k\bar{\ell}}
{}~Y^{\hat{K}}~X^{\hat{L}} \\
{}~~~~=~\left(\hat{R}_{q}^{-1}\right)^{ar}_{~iu}~Y^{i}_{j}
\left(\hat{R}_{q}^{-1}\right)^{ju}_{~ks'}~X^{k}_{\ell'}
 \end{array}  $$

\noindent
Comparing two sides of (5.13) we obtain (5.11). From (5.7) and (4.42)
we have:
$$\begin{array}{l}
\{ig/\lambda\}~\left( Y^{0}X^{I}~-~(-1)^{n_{x}n_{y}}~X^{I}Y^{0}\right)~
\left(\sigma_{I}\right)^{a}_{~b}\\
{}~~~~=~\left(\sigma_{I}\right)^{a}_{~b}
{}~{\bf C}_{\hat{J}\hat{K}}^{~I}~Y^{\hat{J}}X^{\hat{K}}\\
{}~~~~=~\left(-ig q^{1-N+2/N} [N]^{-1/2}\right)^{-1}~
\chi_{\hat{I}}\left(T^{a}_{~b}\right)~{\bf C}_{\hat{J}\hat{K}}^{~\hat{I}}
{}~Y^{\hat{J}}X^{\hat{K}}\\
{}~~~~=~\left(-ig q^{1-N+2/N} [N]^{-1/2}\right)^{-1}~
\left\{Y^{\hat{J}}~\chi_{\hat{J}}\left(T^{a}_{~d}\right)~
X^{\hat{K}}~\chi_{\hat{K}}\left(T^{d}_{~b}\right)\right. \\
{}~~~~~~~\left.~-~
(-1)^{n_{x}n_{y}}~X^{\hat{J}}~\chi_{\hat{J}}\left(T^{a}_{~d}\right)~
Y^{\hat{K}}~\chi_{\hat{K}}\left(T^{d}_{~b}\right) \right\} \\
{}~~~~=~\left(-ig q^{1-N+2/N} [N]^{-1/2}\right)~
\left(Y^{a}_{~d}~X^{d}_{~b}~-~(-1)^{n_{x}n_{y}}~X^{a}_{~d}~
Y^{d}_{~b}\right) \\
{}~~~~~~~~+~\{ig/\lambda\}~(1-q^{2/N})~\left( Y^{0}~X^{a}_{~b}
{}~-~(-1)^{n_{x}n_{y}}~
{}~X^{a}_{~b}~Y^{0}\right)\\
{}~~~~=~\left(-ig q^{1-N} [N]^{-1/2}\right)~
\left(Y^{a}_{~d}~X^{d}_{~b}~-~(-1)^{n_{x}n_{y}}~X^{a}_{~d}~
Y^{d}_{~b}\right)
\end{array} $$

\noindent
(5.12) was proved. $~~~~$ Q.E.D.

\vspace{10mm}
\noindent
{\bf 6. $q$-Deformed Chern Class and $q$-Trace}

In our previous paper [HHM], omitting the possible constant factor,
we assume that the second $q$-deformed Chern class $P$ for the
quantum group $SU_{q}(2)$ has the following form:
$$P~=~\langle F~,~F\rangle~\equiv~F^{I}~F^{J}~g'_{IJ} \eqno (6.1) $$

\noindent
where from the condition:
$$\delta P~=~0,~~~~d P~=~0  \eqno (6.2) $$

\noindent
we defined the $q$-deformed Killing form $g'_{IJ}$ as:
$$g'_{IJ}~=~D^{R}_{~S}~{\bf C}_{IT}^{~S}~{\bf C}_{JR}^{~T}
\eqno (6.3) $$

\noindent
For $SU_{q}(N)$ the non-vanishing components are:
$$\begin{array}{l}
g'_{(k\bar{k})(k\bar{k})}~=~-~g^{2}~q^{2k-2N}~\{[2N]/[N]\}\\
g'_{(j\bar{k})(k\bar{j})}~=~-~g^{2}~q^{j-k-N+2}~\{[2N]/[N]\},~~~~j \neq k
\end{array} $$

\noindent
where the repeated indices are not summed.

Recalling (4.35) and (4.36), we may define another $q$-deformed
Killing form $g_{IJ}$:
$$\begin{array}{l}
g_{IJ}~=~D^{a}_{~b}~\chi_{I}(T^{b}_{~c})~
\chi_{J}(T^{c}_{~a}) \\
g_{(k\bar{k})(k\bar{k})}~=~-~g^{2}~[N]~q^{2k-3N+4/N} \\
g_{(j\bar{k})(k\bar{j})}~=~-~g^{2}~[N]~q^{j-k+2-2N+4/N},~~~~j \neq k
\end{array} \eqno (6.4) $$

\noindent
where the repeated indices are not summed. The else
components of $g_{JK}$ are vanishing. Both for $SU_{q}(2)$ and for
$SU_{q}(N)$ two $q$-deformed Killing forms are
proportional to each other, namely, just like the Killing form
in a Lie algebra, the $q$-deformed Killing form is also independent
of the representation in which it is calculated.

Now, we are going to define the higher $q$-deformed Chern class $P_{m}$
for the quantum group $SU_{q}(N)$ from the covariant condition (6.2).
Generalizing (6.4) we define the "generalized $q$-deformed Killing
forms"and the $q$-trace as follows:
$$g_{I_{1}I_{2}\cdots I_{m}}~=~D^{a_{0}}_{~a_{1}}~
\chi_{I_{1}}\left(T^{a_{1}}_{~a_{2}}\right)~
\chi_{I_{2}}\left(T^{a_{2}}_{~a_{3}}\right)~\cdots~
\chi_{I_{m}}\left(T^{a_{m}}_{~a_{0}}\right)~
\eqno (6.5) $$
$$\langle X_{1}~,~X_{2}~,~\cdots~,~X_{m}\rangle
{}~=~X_{1}^{I_{1}}~X_{2}^{I_{2}}~\cdots ~X_{m}^{I_{m}}
{}~g_{I_{1}I_{2}\cdots I_{m}} \eqno (6.6) $$

\noindent
where $X_{i}$ are fields $\eta$, $d\eta$, $A$ or $dA$ in the BRST algebra
${\cal B}$. In (6.6) the fields can also be replaced by, for
example, $XY^{0}$, $Y^{0}X$ or $F$. From the properties of $q$-Pauli
matrices, the sum of all subscripts of nonvanishing components
$g_{I_{1}\cdots I_{m}}$, as weights, has to be zero. The following
theorem is easily proved from (4.43) and the definition (6.5).

\vspace{5mm}
\noindent
{\bf Proposition 3.} {\it The generalized $q$-deformed Killing forms
satisfy the following relations:}
$$\left({\cal P}_{Adj}\right)^{RS}_{~JK}~
g_{I_{1}\cdots I_{n-1}RSI_{n+2} \cdots I_{m}}
{}~=~-~ig\xi~(\lambda^{2}+2)^{-1}~f_{JK}^{~R}~
g_{I_{1}\cdots I_{n-1}RI_{n+2} \cdots I_{m}}\eqno (6.7) $$

\noindent
where $\xi$ was given in (4.44).

\vspace{3mm}
(6.7) can be rewritten in another form by removing a factor
$f^{P}_{JK}$:
$${\bar f}^{RS}_{K}~g_{I_{1}\cdots I_{n-1}RSI_{n+2} \cdots I_{m}}
{}~=~ig~\xi ~g_{I_{1}\cdots I_{n-1}KI_{n+2} \cdots I_{m}} \eqno (6.8) $$

The $m$-th $q$-deformed Chern class $P_{m}$ for the quantum group
$SU_{q}(N)$ is defined as follows:
$$\begin{array}{rl}
P_{m}&=~\langle F~,~F~,~\cdots~,~F\rangle \\
&=~F^{I_{1}}~F^{I_{2}}~\cdots ~F^{I_{m}}~g_{I_{1}I_{2}\cdots I_{m}}\\
&=~\left(-ig[N]^{-1/2}q^{1-N+2/N}\right)^{m}~
D^{a_{0}}_{~a_{1}}~F^{a_{1}}_{~a_{2}}~F^{a_{2}}_{~a_{3}}~\cdots~
F^{a_{m}}_{~a_{0}}
\end{array} \eqno (6.9) $$

{}From (5.9) and (5.12) we have:
$$\begin{array}{rl}
\delta P_{m}&=~q^{2m/N}~\left(-igq^{1-N}[N]^{-1/2}\right)^{m+1}~\left\{
D^{a_{0}}_{~b}~\eta^{b}_{~a_{1}}~F^{a_{1}}_{~a_{2}}~F^{a_{2}}_{~a_{3}}~\cdots~
F^{a_{m}}_{~a_{0}}\right.\\
&~~~\left.~-~D^{b}_{~a_{1}}~F^{a_{1}}_{~a_{2}}~F^{a_{2}}_{~a_{3}}~\cdots~
F^{a_{m}}_{~a_{0}}~\eta^{a_{0}}_{~b} \right\}
\end{array} \eqno (6.10) $$

By (2.14), (2.15) and (5.11) the second term cancels the first term:
$$\begin{array}{l}
D^{b}_{~a_{1}}~F^{a_{1}}_{~a_{2}}~F^{a_{2}}_{~a_{3}}~\cdots~
F^{a_{m}}_{~a_{0}}~\eta^{a_{0}}_{~b} \\
{}~~~~=~D^{b}_{~a_{1}}~F^{a_{1}}_{~a_{2}}~F^{a_{2}}_{~a_{3}}~\cdots~
F^{a_{m}}_{~a_{0}}\left\{D^{i}_{~j}~q^{N}~\left(\hat{R}_{q}^{-1}
\right)^{a_{0}j}_{~di}\right\}
{}~\eta^{d}_{~b} \\
{}~~~~=~q^{N}~D^{b}_{~a_{1}}~D^{i}_{~j}~F^{a_{1}}_{~a_{2}}~F^{a_{2}}_{~a_{3}}
{}~\cdots~F^{a_{m-1}}_{~a_{m}}~F^{a_{m}}_{~a_{0}}~\left(\hat{R}_{q}^{-1}
\right)^{a_{0}j}_{~dk}~\eta^{d}_{~c}~\left(\hat{R}_{q}^{-1}
\right)^{ck}_{~r\ell}~\left(\hat{R}_{q}\right)^{r\ell}_{~bi} \\
{}~~~~=~q^{N}~D^{b}_{~a_{1}}~D^{i}_{~j}~F^{a_{1}}_{~a_{2}}~F^{a_{2}}_{~a_{3}}
{}~\cdots~F^{a_{m-1}}_{~a_{m}}~\left(\hat{R}_{q}^{-1}
\right)^{a_{m}j}_{~dk}~\eta^{d}_{~c}~\left(\hat{R}_{q}^{-1}
\right)^{ck}_{~r\ell}~F^{r}_{~a_{0}}~\left(\hat{R}_{q}
\right)^{a_{0}\ell}_{~bi} \\
{}~~~~=~q^{N}~D^{b}_{~a_{1}}~D^{i}_{~j}~\left(\hat{R}_{q}^{-1}
\right)^{a_{1}j}_{~dk}~\eta^{d}_{~c}~\left(\hat{R}_{q}^{-1}
\right)^{ck}_{~r\ell}~F^{r}_{~a_{2}}~F^{a_{2}}_{~a_{3}}
{}~\cdots~F^{a_{m-1}}_{~a_{m}}
{}~F^{a_{m}}_{~a_{0}}~\left(\hat{R}_{q}
\right)^{a_{0}\ell}_{~bi} \\
{}~~~~=~q^{N}~D^{a_{0}}_{~a_{1}}~D^{i}_{~j}
{}~\eta^{a_{1}}_{~c}~\left(\hat{R}_{q}^{-1}
\right)^{cj}_{~ri}~F^{r}_{~a_{2}}~F^{a_{2}}_{~a_{3}}
{}~\cdots~F^{a_{m-1}}_{~a_{m}}
{}~F^{a_{m}}_{~a_{0}}\\
{}~~~~=~D^{a_{0}}_{~a_{1}}
{}~\eta^{a_{1}}_{~r}~F^{r}_{~a_{2}}~F^{a_{2}}_{~a_{3}}
{}~\cdots~F^{a_{m-1}}_{~a_{m}}
{}~F^{a_{m}}_{~a_{0}}\end{array} $$

\noindent
Thus, $\delta P_{m}=0$. This technique of proof was firstly used by
Isaev [Isa]. The proof of $d P_{m}=0$ can be performed analogously:
$$\begin{array}{rl}
d P_{m}&=~\left(gq^{1-N+1/N}[N]^{-1/2}\right)^{2m}~\left\{
D^{a_{0}}_{~b}~A^{b}_{~a_{1}}~dA^{a_{1}}_{~a_{2}}~dA^{a_{2}}_{~a_{3}}~\cdots~
dA^{a_{m}}_{~a_{0}}\right.\\
&~~~\left.~-~D^{b}_{~a_{1}}~dA^{a_{1}}_{~a_{2}}~dA^{a_{2}}_{~a_{3}}~\cdots~
dA^{a_{m}}_{~a_{0}}~A^{a_{0}}_{~b} \right\}\\
&=~0
\end{array} $$

Note that the components of the identity and the adjoint representations
are separated in the $q$-deformed Chern class, although they are mixed
in the commutative relations of BRST algebra.

This technique can be used to prove more general relations. Remind
(6.6) and that the condition $\delta P_{m}=0$ can be rewritten as follows:
$$\begin{array}{rl}
\delta P_{m}&=~\{ig/\lambda\}~\left\{\eta^{0}~F^{I_{1}}~F^{I_{2}}
{}~\cdots ~F^{I_{m}}~g_{I_{1}I_{2}\cdots I_{m}}
{}~-~F^{I_{1}}~F^{I_{2}}~\cdots ~F^{I_{m}}~\eta^{0}~
g_{I_{1}I_{2}\cdots I_{m}} \right\}\\
&=~0 \end{array} $$

\noindent
Now, the following theorem can be proved straightforwardly.

\vspace{5mm}
\noindent
{\bf Proposition 4.} {\it Let $Y \in {\cal B}$ with index $n_{y}$, and
$X_{i}$, that are not the field $Y$, be any fields in ${\cal B}$ with
the indices $n_{i}$, $n_{i}> n_{y}$. Then}:
$$\begin{array}{l}
(-1)^{n}~\langle X_{1}~,~\cdots~,~X_{m}~,~Y \rangle~=~
\langle Y~,~X_{1}~,~\cdots~,~X_{m} \rangle \\
(-1)^{n}~\langle X_{1}~,~\cdots~,~X_{m} \rangle~Y^{0}~=~
Y^{0}~\langle X_{1}~,~\cdots~,~X_{m} \rangle \\
n~=~\displaystyle \sum_{i=1}^{m}~n_{i}~n_{y}
\end{array} \eqno (6.11) $$

\vspace{3mm}
Some corollaries can be derived from Proposition 4. First of all,
substituting the commutative relation (5.6) into (6.11) we obtain
some constraints for the "generalized $q$-deformed Killing forms":
$$\begin{array}{l}
g_{I_{1}I_{2}\cdots I_{m}I_{0}}
{}~\Lambda^{I_{1}\hat{K}_{1}}_{~J_{0}J_{1}}~
\Lambda^{I_{2}\hat{K}_{2}}_{~\hat{K}_{1}J_{2}}~\cdots
{}~\Lambda^{I_{m-1}\hat{K}_{m-1}}_{~\hat{K}_{m-2}J_{m-1}}~
\Lambda^{I_{m}I_{0}}_{~\hat{K}_{m-1}J_{m}} ~=~
g_{J_{0}J_{1}\cdots J_{m}} \\
g_{I_{1}I_{2}\cdots I_{m}I_{0}}
{}~\Lambda^{I_{1}\hat{K}_{1}}_{~0J_{1}}~
\Lambda^{I_{2}\hat{K}_{2}}_{~\hat{K}_{1}J_{2}}~\cdots
{}~\Lambda^{I_{m-1}\hat{K}_{m-1}}_{~\hat{K}_{m-2}J_{m-1}}~
\Lambda^{I_{m}I_{0}}_{~\hat{K}_{m-1}J_{m}}~=~0 \end{array}
\eqno (6.12) $$
$$g_{I_{1}I_{2}\cdots I_{m}}
{}~\Lambda^{I_{1}\hat{K}_{1}}_{~\hat{K}_{0}J_{1}}~
\Lambda^{I_{2}\hat{K}_{2}}_{~\hat{K}_{1}J_{2}}~\cdots
{}~\Lambda^{I_{m-1}\hat{K}_{m-1}}_{~\hat{K}_{m-2}J_{m-1}}~
\Lambda^{I_{m}0}_{~\hat{K}_{m-1}J_{m}}
{}~=~\delta^{0}_{\hat{K}_{0}}~g_{J_{1}J_{2}\cdots J_{m}}  \eqno (6.13) $$

\noindent
(6.12) describes the cyclic property of the $q$-trace.
(6.13) is equivalent to the following form:
$$\begin{array}{l}
\displaystyle \sum_{n=1}^{m}~
g_{I_{1}\cdots I_{n}J_{n+1}\cdots J_{m}}
{}~\Lambda^{I_{1}\hat{K}_{1}}_{~\hat{K}_{0}J_{1}}~
\Lambda^{I_{2}\hat{K}_{2}}_{~\hat{K}_{1}J_{2}}~\cdots
{}~\Lambda^{I_{n-1}\hat{K}_{n-1}}_{~\hat{K}_{n-2}J_{n-1}}~
{\bf C}^{~I_{n}}_{\hat{K}_{n-1}J_{n}}~=~0
\end{array} \eqno (6.14) $$

\noindent
namely,
$$\begin{array}{l}
g_{I_{1}I_{2}\cdots I_{m}}
{}~\chi_{\hat{K}_{0}} \left(M^{I_{1}}_{~J_{1}}\cdots M^{I_{m}}_{~J_{m}}\right)
{}~=~0 \end{array} \eqno (6.15) $$

\noindent
Bernard [Ber] gave the special form ($m=2$) of (6.14):
$$\begin{array}{l}
g_{I_{1}J_{2}}~{\bf C}^{~I_{1}}_{\hat{K}_{0}\hat{J}_{1}}
{}~+~g_{I_{1}I_{2}}~\Lambda^{I_{1}\hat{K}_{1}}_{~\hat{K}_{0}\hat{J}_{1}}~
{\bf C}^{~I_{2}}_{\hat{K}_{1}\hat{J}_{2}}~=~0
\end{array}\eqno (6.16) $$

In terms of Proposition 3, it can be proved by direct calculation that
the constraint (6.12) is equivalent to (6.13). It is not surprising
because they come from the same source. In fact, by making use
of the explicit form (4.24) we obtain the same relations
from (6.13) as from (6.12). For example, when $m=1$ we obtain:
$$g_{I}~=~0 \eqno (6.17) $$

\noindent
It implies the orthogonality (3.9) of $q$-Pauli matrices. When $m=2$,
we obtain:
$$\begin{array}{l}
(\lambda^{2}+2)~g_{IJ}~+~g_{RS}~\bar{f}^{RT}_{I}~f^{S}_{TJ}~=~0\\
(\lambda^{2}+1)~g_{RK}~f^{R}_{IJ}~+~g_{IR}~f^{R}_{JK}~+
g_{RS}~\bar{f}^{RT}_{P}~f^{P}_{IJ}~f^{S}_{TK}~=~0\end{array} $$

\noindent
Thus,
$$ f_{IJ}^{R}~g_{RK}~=~g_{IR}~f_{JK}^{R}\eqno (6.18) $$

\noindent
When $m=3$ we have:
$$\begin{array}{l}
(\lambda^{4}+3\lambda^{2}+3)~g_{IJK}~+~(\lambda^{2}+1)~\left\{
\bar{f}^{RT}_{I}~f^{S}_{TJ}~g_{RSK}~+~\bar{f}^{RT}_{J}~f^{S}_{TK}~g_{IRS}
\right\} \\
{}~~~~+~\bar{f}^{RS}_{I}~f^{T}_{JK}~g_{RST}~+~\bar{f}^{RP_{3}}_{I}
{}~f^{P_{2}}_{P_{3}J}~\bar{f}^{SP_{1}}_{P_{2}}~f^{T}_{P_{1}K}~g_{RST}~=~0 \\
(\lambda^{2}+2)~f^{R}_{LI}~g_{RJK}~-~
(\lambda^{2}+1)~f^{R}_{IJ}~g_{LRK}~-~
{}~f^{R}_{JK}~g_{LIR} \\
{}~~~~=~\bar{f}^{RT}_{P}~f^{P}_{IJ}~f^{S}_{TK}~g_{LRS}
{}~-~f^{R}_{LI}~\bar{f}^{SP}_{J}~f^{T}_{PK}~g_{RST}
\end{array} \eqno (6.19) $$

At last, from Propositions 3 and 4 we can prove:
$$\begin{array}{c}
\langle A^{m} \rangle~\equiv~
A^{I_{1}}~A^{I_{2}}~\cdots~A^{I_{m}}~g_{I_{1}I_{2}\cdots I_{m}}~=~0
,~~~~{\rm when}~~m\neq 3 \\
\end{array} \eqno (6.20) $$

\noindent
In fact, define:
$$\begin{array}{rl}
\langle A^{m} \rangle~\phi_{n}&=~
g_{T_{1}\cdots T_{m-n}I_{1}\cdots I_{n}}
{}~\Lambda^{I_{1}\hat{K}_{1}}_{~J_{1}J_{2}}
\Lambda^{I_{2}\hat{K}_{2}}_{~\hat{K}_{1}J_{3}}~\cdots
\Lambda^{I_{n-2}\hat{K}_{n-2}}_{~\hat{K}_{n-3}J_{n-1}}
\Lambda^{I_{n-1}I_{n}}_{~\hat{K}_{n-2}J_{n}} \\
&~~~\cdot~A^{T_{1}}\cdots A^{T_{m-n}}A^{J_{1}}\cdots A^{J_{n}},~~~~
2\leq n\leq m \\
\langle A^{m} \rangle~\psi_{n}&=~
g_{T_{1}\cdots T_{m-n}I_{1}\cdots I_{n}}
{}~\Lambda^{I_{1}0}_{~J_{1}J_{2}}
\Lambda^{I_{2}\hat{K}_{2}}_{~0J_{3}}~\cdots
\Lambda^{I_{n-2}\hat{K}_{n-2}}_{~\hat{K}_{n-3}J_{n-1}}
\Lambda^{I_{n-1}I_{n}}_{~\hat{K}_{n-2}J_{n}} \\
&~~~\cdot~A^{T_{1}}\cdots A^{T_{m-n}}A^{J_{1}}\cdots A^{J_{n}},~~~~
3\leq n\leq m \end{array} \eqno (6.21) $$

\noindent
Since the factor $\langle A^{m} \rangle$ has been abstracted, the functions
$\phi_{n}$and $\psi_{n}$ are independent of $m$. From (4.24), (4.27) and
(5.3) we obtain the recurrence relations and some explicit values for
the functions:
$$\begin{array}{l}
\phi_{n}~=~-~(\lambda^{2}+1)~\phi_{n-1}~+~\psi_{n},~~~~n\geq 3\\
\psi_{n}~=~(\lambda^{2}+1)\psi_{n-1}~-~\lambda^{2}(\lambda^{2}+1)\psi_{n-2} \\
{}~~~~~~~~+~\displaystyle \sum_{a=1}^{n-5} ~\left\{(-1)^{a+1}\lambda^{2}
\left((\lambda^{2}+1)^{2}-a(\lambda^{2}+2)\right)\psi_{n-a-2}\right\}\\
{}~~~~~~~~+~(-1)^{n}\lambda^{2}(\lambda^{2}+2)(\lambda^{2}+1)^{n-4}
(\lambda^{2}-n+4),~~~~n \geq 5 \end{array} $$
$$\begin{array}{l}
\phi_{2}~=~-\lambda^{2}-1,~~~~ \phi_{3}~=~1,~~~~\phi_{4}~=~
-~\lambda^{2}(\lambda^{2}+3)~-~1\\
\phi_{5}~=~\lambda^{2}(\lambda^{2}+2)^{2}+1,~~~~
\phi_{6}~=~-~\lambda^{2}(\lambda^{2}+3)(\lambda^{4}+3\lambda^{2}+3)~-~1
\end{array} $$

The limit values of the functions are as follows:
$$\begin{array}{l}
\displaystyle \lim_{q\rightarrow 1}~\phi_{2n}~=~-~1 ,~~~~
\displaystyle \lim_{q\rightarrow 1}~\phi_{2n+1}~=~1 ,~~~~
\displaystyle \lim_{q\rightarrow 1}~\psi_{n}~=~0 \\
\displaystyle \lim_{q\rightarrow 1}~\lambda^{-2}~\left(\phi_{2n}+1\right)
{}~=~-~2n^{2}+4n-3\\
\displaystyle \lim_{q\rightarrow 1}~\lambda^{-2}~\left(\phi_{4n+1}-1\right)
{}~=~4n(2n-1),~~~~
\displaystyle \lim_{q\rightarrow 1}~\lambda^{-2}~\left(\phi_{4n+3}+1\right)
{}~=~4n(2n+1)\\
\displaystyle \lim_{q\rightarrow 1}~\lambda^{-2}~\psi_{2n}
{}~=~-~2(n-1),~~~~
\displaystyle \lim_{q\rightarrow 1}~\lambda^{-2}~\psi_{2n+1}
{}~=~2(n-2) \end{array} $$

Now, substituting (6.12) into (6.21) we have:
$$\left(\phi_{m}-1\right)~\langle A^{m} \rangle~=~0 \eqno (6.22)$$

\noindent
It leads to (6.20). It is interesting to notice that in the classical
case ($q=1$) only $\langle A^{2n} \rangle=0$ is well known. (6.20) also
holds for $\eta$ instead of $A$ due to (4.32).

Generalize the $q$-trace (6.6) as follows:
$$\begin{array}{l}
\langle \cdots~,~Z_{1}~,~[X~,~Y]~,~Z_{2}~,~\cdots\rangle
{}~=~\cdots Z_{1}^{I_{1}}~X^{J}~Y^{K}~Z_{2}^{I_{2}}\cdots
\left({\cal P}_{Adj}\right)^{RS}_{~JK}~g_{\cdots I_{1}RSI_{2}\cdots }
\end{array} \eqno (6.23) $$

\noindent
where $X$, $Y$ and $Z$ are fields in the BRST algebra
${\cal B}$. The fields can also be replaced by, for
example, $XY^{0}$, $Y^{0}X$ or $F$. From Propositions 3 and 4,
we have:
$$\begin{array}{l}
\langle \cdots~,~X,~[ \eta~,~ \eta],~Y,~\cdots \rangle
{}~=~\langle \cdots~,~X,~ \eta~,~ \eta,~Y,~\cdots \rangle \\
\langle \cdots~,~X,~[ A~,~ A],~Y,~\cdots \rangle
{}~=~\langle \cdots~,~X,~ A~,~ A,~Y,~\cdots \rangle \\
\langle ~\cdots,~Z_{1},~[X~,~Y],~Z_{2}~\cdots~ \rangle\\
{}~~~~~~~~=~\cdots~
Z_{1}^{I_{1}}\left\{-ig \xi (\lambda^{2}+2 )^{-1}~ f_{JK}^{T}
X^{J}Y^{K}\right\}~ Z_{2}^{I_{2}}~\cdots~
g_{\cdots I_{1}TI_{2} \cdots}
\end{array} \eqno (6.24) $$
$$\begin{array}{l}
\left({\cal P}_{Adj}\right)^{RS}_{~JK}~g_{RS}~=~0 ,~~~~
\bar{f}_{J}^{RS}~g_{RS}~=~0,~~~~
\Lambda^{RS}_{~JK}~g_{RS}~=~g_{JK} \end{array} \eqno (6.25) $$
$$\begin{array}{l}
\left({\cal P}_{Adj}\right)^{RS}_{~IJ}~g_{RSK}~=~
g_{IRS}~\left({\cal P}_{Adj}\right)^{RS}_{~JK},~~~~
\langle ~X,~[ Y~,~Z]  \rangle
{}~=~\langle [X~,~Y],~Z \rangle
\end{array} \eqno (6.26) $$

\noindent
where $X$, $Y$ and $Z$ are fields in the BRST algebra
${\cal B}$.

\vspace{5mm}
\noindent
{\bf Proposition 5.} {\it Assume $X^{J}$ and $Y^{J}$ are two different
fields in ${\cal B}$ with the indices $n_{x}>n_{y}$. $Z^{J}$ denotes a
field in ${\cal B}$, or a field multiplying by the zero component
of this or another field. Since the following relations are
linear for the field $Z^{J}$, $Z^{J}$ can also be replaced by their
linear combination, for example, $F^{J}$. Then,}
$$\begin{array}{l}
(-1)^{n_{x}n_{y}}~\langle \cdots,~Z_{1}~,~XY^{0}~,~Z_{2}~,\cdots \rangle \\
{}~~~~=(\lambda^{2}+1)\langle \cdots,Z_{1},Y^{0}X,Z_{2},\cdots \rangle
{}~-~\displaystyle {\lambda(\lambda^{2}+2) \over
ig \xi}\langle \cdots, Z_{1},[Y,X] ,Z_{2},\cdots \rangle
\end{array} \eqno (6.27) $$
$$\begin{array}{l}
\{ig /\lambda\}~\langle \cdots~,~Z_{1}~,~(A^{0}A+AA^{0})~,~Z_{2}
{}~,~\cdots \rangle \\
{}~~~~~~=~\xi^{-1}~\langle \cdots~,~Z_{1}~,~A~,~A~,~Z_{2}~,~\cdots \rangle \\
\{ig /\lambda\}~\langle \cdots~,~Z_{1}~,~(\eta^{0}\eta+\eta \eta^{0})~,~Z_{2}
{}~,~\cdots \rangle \\
{}~~~~~~=~\xi^{-1}~\langle \cdots~,~Z_{1}~,~\eta~,~\eta~,~Z_{2}~,~\cdots
\rangle
\end{array} \eqno (6.28) $$
$$\begin{array}{l}
(-1)^{n_{x}n_{y}}~\langle \cdots~,~Z_{1}~,~X~,~Y~,~Z_{2}
{}~,~\cdots \rangle\\
{}~~~~=~\langle \cdots~,~Z_{1}~,~Y~,~X~,~Z_{2}~,~\cdots \rangle \\
{}~~~~~~~~-~ig\lambda^{-1}\xi~
\langle \cdots~,~Z_{1}~,~(Y^{0}X-(-1)^{n_{x}n_{y}}XY^{0})~,~Z_{2}~,~
\cdots \rangle \end{array} \eqno (6.29) $$
$$\begin{array}{l}
(-1)^{n_{x}n_{y}}~\langle \cdots,~Z_{1}~,~X~,~Y ~,~Z_{2}~,\cdots \rangle\\
{}~~~~=~ig\lambda \xi~\langle \cdots,~Z_{1}~,~Y^{0}X
{}~,~Z_{2}~,\cdots \rangle \\
{}~~~~~~~~-~(\lambda^{2}+2)~
\langle \cdots,~Z_{1}~,~[Y~,~X] ~,~Z_{2}~,\cdots \rangle
{}~+~ \langle \cdots,~Z_{1}~,~Y~,~X ~,~Z_{2}~,\cdots \rangle  \\
(-1)^{n_{x}n_{y}}~\langle \cdots,~Z_{1}~,~[X~,~Y] ~,~Z_{2}~,\cdots \rangle\\
{}~~~~=~ig\lambda \xi~\langle \cdots,Z_{1},Y^{0}X
,Z_{2},\cdots \rangle ~-~(\lambda^{2}+1)
\langle \cdots,Z_{1},[Y,X] ,Z_{2},\cdots \rangle
\end{array} \eqno (6.30) $$

\noindent
{\it Proof.} From (5.7) and (4.36) we have
$$(-1)^{n_{x}n_{y}}~X^{I}~Y^{0}~=~(\lambda^{2}+1)~Y^{0}~X^{I}~+~
{}~\lambda~Y^{J}~X^{K}~f_{JK}^{I}$$

\noindent
Due to (4.43),
$$\begin{array}{l}
(-1)^{n_{x}n_{y}}~X^{I}Y^{0}~\chi_{I}\left(T^{a}_{~b}\right)
{}~=~(\lambda^{2}+1)~Y^{0}X^{I}~\chi_{I}\left(T^{a}_{~b}\right) \\
{}~~~~~~~~-~\displaystyle {\lambda(\lambda^{2}+2) \over ig\xi}~
Y^{I}X^{L}~\left({\cal P}_{Adj}\right)^{JK}_{~IL}
\chi_{J}\left(T^{a}_{~d}\right)\chi_{K}\left(T^{d}_{~b}\right)
\end{array} $$

\noindent
According to the definition (6.6) we obtain (6.27).

{}From (5.3) we have:
$$\begin{array}{rl}
\{ig /\lambda\}~\left( A^{0}A^{I}~+~A^{I}A^{0}\right)~
\chi_{I}\left(T^{a}_{~b}\right)
&=~(\lambda^{2}+2)^{-1}~A^{J}A^{K}~
{\bf C}_{JK}^{~I}\chi_{I}\left(T^{a}_{~b}\right)\\
&=~\xi^{-1}~A^{J}A^{K}~\chi_{J}\left( T^{a}_{~d}\right)\chi_{K}\left(
T^{d}_{~b}\right)
\end{array}  $$

\noindent
This relation leads to (6.28). The relation also holds for $\eta^{J}$
due to (4.32).

{}From (4.42) and (5.7) we obtain:
$$\begin{array}{l}
(-1)^{n_{x}n_{y}}~X^{\hat{I}}Y^{\hat{J}}~\chi_{\hat{I}}\left(T^{a}_{~d}
\right)~\chi_{\hat{J}}\left(T^{d}_{~b}\right)
{}~=~Y^{\hat{K}}X^{\hat{L}}~\Lambda^{\hat{I}\hat{J}}_{~\hat{K}\hat{L}}
{}~\chi_{\hat{I}}\left(T^{a}_{~d}\right)~\chi_{\hat{J}}\left(T^{d}_{~b}\right)\\
{}~~~~=~Y^{\hat{K}}X^{\hat{L}}~\left\{~\chi_{\hat{K}}\left(T^{a}_{~d}
\right)~\chi_{\hat{L}}\left(T^{d}_{~b}\right)~-~{\bf C}_{\hat{K}\hat{L}}^{
{}~T}\chi_{T}\left(T^{a}_{~b}\right) \right\} \\
{}~~~~=~Y^{\hat{K}}X^{\hat{L}}~\chi_{\hat{K}}\left(T^{a}_{~d}
\right)~\chi_{\hat{L}}\left(T^{d}_{~b}\right)
{}~-~\{ig/\lambda\}~\left(Y^{0}X^{J}~-~(-1)^{n_{x}n_{y}}~X^{J}Y^{0}\right)~
\chi_{J}\left(T^{a}_{~b}\right)  \end{array} $$

\noindent
Then, noting (4.35) and (5.7) we have:
$$\begin{array}{l}
\left\{(-1)^{n_{x}n_{y}}~X^{I}Y^{J}~-~Y^{I}X^{J}\right\}
{}~\chi_{I}\left(T^{a}_{~d}\right)~\chi_{J}\left(T^{d}_{~b}\right)\\
{}~~~~=~\left\{(-1)^{n_{x}n_{y}}~X^{\hat{I}}Y^{\hat{J}}~-~
Y^{\hat{I}}X^{\hat{J}}\right\}
{}~\chi_{\hat{I}}\left(T^{a}_{~d}\right)~\chi_{\hat{J}}\left(T^{d}_{~b}\right)\\
{}~~~~~~~~-~ig~\left\{q^{-N+2/N}[N]^{-1}+ \lambda^{-1}\left(1-q^{2/N}\right)
\right\}\left\{(-1)^{n_{x}n_{y}}X^{J}Y^{0}-Y^{0}X^{J}\right\}
\chi_{J}\left(T^{a}_{~b}\right) \\
{}~~~~=~-~ig\lambda^{-1}\xi~
\left\{Y^{0}X^{J}~-~(-1)^{n_{x}n_{y}}~X^{J}Y^{0}\right\}~
\chi_{J}\left(T^{a}_{~b}\right) \end{array} $$

\noindent
It leads to (6.29). From (6.27) and (6.29) we obtain (6.30). $~~~~$ Q.E.D.

{}From Proposition 5 we obtain:
$$\begin{array}{rl}
\langle F~,~ A \rangle&=~\langle A~,~ F \rangle~=~\langle A~,~ dA \rangle
{}~+~\{ig / \lambda \}~\langle A~,~(A^{0}A+AA^{0}) \rangle\\
&=~\langle A~,~ dA \rangle ~+~\xi^{-1}
{}~\langle A~,~A~,~A \rangle
\end{array}\eqno (6.31) $$

\vspace{5mm}
\noindent
{\bf Proposition 6.} {\it For any field $X \in {\cal B}$,}
$$\langle \eta~,~\eta~,~[\eta~,~X] \rangle~=~0,~~~~
\langle A~,~A~,~[A~,~X] \rangle~=~0 \eqno (6.32) $$

\noindent
{\it Proof.} From (4.27) and (4.32) we have:
$$\begin{array}{rl}
f_{IL}^{T}~f_{JK}^{L}~\eta^{I}~\eta^{J}~X^{K}
&=~-~(\lambda^{2}+2)^{-1}~f_{IL}^{T}~f_{JK}^{L}~
\bar{f}^{IJ}_{P}~f_{RS}^{P}~\eta^{R}~\eta^{S}~X^{K}\\
&=~\displaystyle {\lambda^{2}+1 \over \lambda^{2}+2 }~
f_{RS}^{P}~f_{PK}^{T}~\eta^{R}~\eta^{S}~X^{K} \end{array} \eqno (6.33) $$

\noindent
Noting (6.7) and (6.18) we have:
$$\begin{array}{rl}
\left(\displaystyle {\lambda^{2}+2 \over -ig\xi }\right)^{2}
{}~\langle \eta~,~\eta~,~[\eta~,~X] \rangle&=~\eta^{I}~\eta^{J}~\eta^{K}~X^{L}
{}~f_{IJ}^{T}~f_{KL}^{Q}~g_{TQ}\\
&=~\eta^{I}~\eta^{J}~\eta^{K}~X^{L}
{}~f_{IJ}^{T}~f_{TK}^{Q}~g_{QL}\\
&=~\displaystyle {\lambda^{2}+2 \over \lambda^{2}+1 }~
\eta^{I}~\eta^{J}~\eta^{K}~X^{L}
{}~f_{IT}^{Q}~f_{JK}^{T}~g_{QL}\\
&=~\displaystyle {\lambda^{2}+2 \over \lambda^{2}+1 }~
\eta^{I}~\eta^{J}~\eta^{K}~X^{L}
{}~f_{TL}^{Q}~f_{JK}^{T}~g_{IQ}\\
&=~\left(\displaystyle {\lambda^{2}+2 \over \lambda^{2}+1 }\right)^{2}~
\eta^{I}~\eta^{J}~\eta^{K}~X^{L}
{}~f_{JT}^{Q}~f_{KL}^{T}~g_{IQ}\\
&=~\left(\displaystyle {\lambda^{2}+2 \over \lambda^{2}+1 }\right)^{2}~
\eta^{I}~\eta^{J}~\eta^{K}~X^{L}
{}~f_{IJ}^{Q}~f_{KL}^{T}~g_{QT} \end{array} $$

\noindent
Thus,
$$\left\{1~-~\left(\displaystyle {\lambda^{2}+2 \over
\lambda^{2}+1 }\right)^{2}~
\right\}~\eta^{I}~\eta^{J}~\eta^{K}~X^{L}
{}~f_{IJ}^{T}~f_{KL}^{Q}~g_{TQ}=0 $$

\noindent
The first relation in (6.32) is now proved. The proof for the next
relation can be performed analogously. $~~~~$ Q.E.D.

Let $Z^{J}$ be a field in ${\cal B}$ with the index $n_{z}$, and
let $X^{J}$ denote the field $\eta^{J}$ or $A^{J}$ with the index
$n_{x}=-1$ or $1$, respectively.  From Propositions 4 and 5
we obtain that if $n_{z}<n_{x}$:
$$\begin{array}{l}
\langle X~,~X~,~Z \rangle ~=~\langle Z~,~X~,~X \rangle \\
(-1)^{n_{x}n_{z}}~\langle X~,~Z~,~X \rangle
{}~=~-~(\lambda^{2}+1)~\langle Z~,~X~,~X \rangle \\
\langle X~,~X~,~Z \rangle~-~(-1)^{n_{x}n_{z}} \langle X~,~Z~,~X \rangle
{}~+~\langle Z~,~X~,~X \rangle \\
{}~~~~=~[3]~\langle Z~,~X~,~X \rangle
\end{array} \eqno (6.34) $$

\noindent
If $n_{z}>n_{x}$, we have:
$$\begin{array}{l}
\langle Z~,~X~,~X \rangle~=~ig\lambda \xi(\lambda^{2}+2)~X^{0}
\langle X~,~Z \rangle ~+(\lambda^{2}+1)~\langle X~,~X~,~Z \rangle\\
(-1)^{n_{x}n_{z}}~\langle X~,~Z~,~X \rangle~=~-~ig\lambda \xi~X^{0}
\langle X~,~Z \rangle ~-~\langle X~,~X~,~Z \rangle\\
{}~~~~~~~~=~-~(\lambda^{2}+2)^{-1}~\left\{ \langle X~,~X~,~Z \rangle
{}~+~\langle Z~,~X~,~X \rangle  \right\} \\
\langle X~,~X~,~Z \rangle~-~(-1)^{n_{x}n_{z}} \langle X~,~Z~,~X \rangle
{}~+~\langle Z~,~X~,~X \rangle \\
{}~~~~~~~~=~-~(-1)^{n_{x}n_{z}}~[3]~\langle X~,~Z~,~X \rangle
\end{array} \eqno (6.35) $$

\vspace{10mm}
\noindent
{\bf 7. $q$-Deformed Chern-Simons and Cocycle Hierarchy}

In the classical case Zumino [Zum1] [MSZ] introduced a homotopy operator
$k$ to compute the Chern-Simons. Generalizing his method we compute
the second $q$-deformed Chern-Simons for $SU_{q}(2)$ in our previous
paper [HHM]. Now, we compute the $m$-th $q$-deformed Chern-Simons
for $SU_{q}(N)$.

Introduce a $q$-deformed homotopy operator $k$ that is nilpotent and
satisfies the Leibniz rule in the graded sense for the index $n$:
$$k^{2}~=~0,~~~~d~k~+~k~d~=~{\bf 1} \eqno (7.1) $$

\noindent
In the following we are going to show the existence of $k$, and
compute the $q$-deformed Chern-Simons $Q_{2m-1}(A)$ from the $m$-th
$q$-deformed Chern class by the operator $k$:
$$\begin{array}{c}
P_{m}~=~(d~k~+~k~d)~P_{m}~=~d\left(k~P_{m} \right)~=~d~Q_{2m-1}(A) \\
Q_{2m-1}(A)~=~k~P_{m} \end{array} \eqno (7.2) $$

\noindent
where we used $dP_{m}=0$.

Introduce a real parameter $t$, $0\leq t \leq 1$. When $t$ changes
from $0$ to $1$, the gauge potentials $A^{J}_{t}$ change from $0$ to
$A^{J}$:
$$\begin{array}{rl}
A^{J}_{t}&=~t~A^{J}\\
F_{t}^{J}&=~t~dA^{J}~+~\{ig~t^{2}/\lambda\}~
\left(A^{0}~A^{J}~+~A^{J}~A^{0}\right)\\
&=~t~F^{J}~+~\{ig~(t^{2}-t)/ \lambda \}~
\left(A^{0}~A^{J}~+~A^{J}~A^{0}\right)
\end{array} \eqno (7.3) $$

We choose the symmetrized definition for the $q$-deformed derivative
and the $q$-deformed integral [GR] [Jackson] to fit our definition
(2.9) for $q$-number. Define the $q$-deformed derivative along $t$ by:
$$\displaystyle {\partial \over \partial_{q} t}~f(t)
{}~=~\displaystyle {f(qt)~-f(q^{-1}t) \over t~(q~-~q^{-1})} \eqno (7.4)$$

\noindent
satisfying the $q$-deformed Leibniz rule:
$$\displaystyle {\partial \over \partial_{q} t}~f(t)g(t)~=~
\displaystyle {\partial f(t)\over \partial_{q} t}~g(qt)~+~
f(q^{-1}t)~\displaystyle {\partial g(t) \over \partial_{q} t}
\eqno (7.5) $$

The $q$-deformed integral is defined by:
$$\displaystyle \int_{0}^{t_{0}}~d_{q}t~f(t)~=~t_{0}(1-q^{2})~
\displaystyle \sum_{k=0}^{\infty}~q^{2k}~f(q^{2k+1}t_{0}) \eqno (7.6) $$

\noindent
At least for a polynomial, the $q$-deformed integral is the inverse
of $q$-deformed derivative. For example,
$$\displaystyle {\partial \over \partial_{q} t}~t^{m}~=~[m]~t^{m-1},~~~~
\displaystyle \int_{0}^{t_{0}}~d_{q}t~t^{m-1}~=~t_{0}^{m}/[m]$$

Now, define the $q$-deformed Lie derivative $\hat{\delta}_{q}$
along $t$ in the gauge space:
$$\hat{\delta}_{q}~\equiv~d_{q}t~\displaystyle {\partial
\over \partial_{q} t} \eqno (7.7) $$

\noindent
and the $q$-deformed operator $\ell_{t}$ that satisfies the
$q$-deformed Leibniz rule in the graded sense for the index $n$:
$$\begin{array}{l}
\ell_{t} A_{t}^{J}~=~0,~~~~
\ell_{t} F_{t}^{J}~=~\hat{\delta}_{q} A^{J}_{t}~=~d_{q}t~A^{J} \\
\ell_{t}\left\{X_{t}^{J}~Y_{t}^{K} \right\}
{}~=~\left\{\ell_{t} X_{t}^{J} \right\}~Y_{qt}^{K}~+~(-1)^{n}~X_{q^{-1}t}^{J}
{}~\left\{\ell_{t} Y_{t}^{K} )\right\}
\end{array} \eqno (7.8) $$

\noindent
where $X^{J}_{t}$ and $Y^{K}_{t}$ are the fields in ${\cal B}$,
and $X_{t}^{J}$ has the index $n$.

It is easy to check that for all formal polynomials (vanishing at
$F^{J}_{t}=0$ and $A^{J}_{t}=0$) we have
$$\begin{array}{c}
\ell_{t}~\ell_{t}~=~0\\
\ell_{t}~d~+~d~\ell_{t}~=~\hat{\delta}_{q}~=~d_{q}t~\displaystyle {\partial
\over \partial_{q} t} \\
\hat{\delta}_{q}~d~=~d~\hat{\delta}_{q},~~~~
\hat{\delta}_{q}~\ell_{t}~=~\ell_{t}~\hat{\delta}_{q}~
\end{array} \eqno (7.9) $$

\noindent
Comparing it with (7.1) we obtain:
$$k~=~\displaystyle \int_{0}^{1}~\ell_{t} \eqno (7.10) $$

The $(2m-1)$-th $q$-deformed Chern-Simons can be computed from (7.2)
straightforwardly. In the following we give some examples. For $m=2$
we have:
$$\begin{array}{rl}
\ell_{t} \left(P_{2}\right)_{t}&=~\langle \ell_{t}F_{t}~,~F_{qt}\rangle
{}~+~\langle F_{q^{-1}t}~,~\ell_{t}F_{t}\rangle \\
&=~d_{q}t~\left\{ \langle A~,~F_{qt}\rangle~+~\langle
F_{q^{-1}t}~,~A \rangle \right\}\\
&=~d_{q}t~\left\{t[2] \langle A~,~dA \rangle~+~ig\lambda^{-1}t^{2}
(\lambda^{2}+2)~\langle A~,~(A^{0}A+AA^{0}) \rangle \right\}\\
&=~d_{q}t~\left\{ t~[2]~\langle A~,~dA \rangle~+~t^{2}\xi^{-1}
(\lambda^{2}+2) ~\langle A~,~A~,~A \rangle
\right\} \end{array} \eqno (7.11) $$

\noindent
where we have used (6.28).
$$\begin{array}{rl}
Q_{3}(A)&=~k~\left(P_{2}\right)_{t} \\
&=~\langle A~,~dA \rangle~+~ \xi^{-1}\{ [4]/[3][2]\}
{}~\langle A~,~A~,~A\rangle \\
&=~\langle A~,~F\rangle~-~\{\xi [3]\}^{-1}
{}~\langle A~,~A~,~A\rangle \end{array} \eqno (7.12) $$

\noindent
For $m=3$ we have:
$$\begin{array}{rl}
\ell_{t} \left(P_{3}\right)_{t}&=~
\langle \ell_{t}F_{t}~,~F_{qt}~,~F_{qt}\rangle ~+~
\langle F_{q^{-1}t}~,~\ell_{t}F_{t}~,~F_{qt}\rangle ~+~
\langle F_{q^{-1}t}~,~F_{q^{-1}t}~,~\ell_{t}F_{t}\rangle \\
&=~d_{q}t~\langle A~,~F_{qt}~,~F_{qt}\rangle ~+~
\langle F_{q^{-1}t}~,~A~,~F_{qt}\rangle ~+~
\langle F_{q^{-1}t}~,~F_{q^{-1}t}~,~A \rangle \\
&=~d_{q}t~t^{2}~\left\{q^{2}~\langle A~,~dA~,~dA \rangle~+~
\langle dA~,~A~,~dA \rangle~+~q^{-2}~\langle dA~,~dA~,~A \rangle \right\} \\
&~~~~+~d_{q}t~t^{3}~\xi^{-1}~\left\{(q^{3}+q^{-1})~
\langle A~,~A~,~A~,~dA \rangle~+~q^{-3}~\langle A~,~A~,~dA~,~A \rangle
\right.\\
&\left.~~~~+~q^{3}~\langle A~,~dA~,~A~,~A \rangle
{}~+~(q+q^{-3})~\langle dA~,~A~,~A~,~A \rangle \right\}
\end{array} $$
$$\begin{array}{rl}
Q_{5}(A)&=~k~\left(P_{3}\right)_{t} \\
&=~[3]^{-1}~\left\{q^{2}~\langle A~,~dA~,~dA \rangle~+~
\langle dA~,~A~,~dA \rangle~+~q^{-2}~\langle dA~,~dA~,~A \rangle \right\} \\
&~~~~+~\{\xi [4]\}^{-1}~\left\{(q^{3}+q^{-1})~
\langle A~,~A~,~A~,~dA \rangle~+~q^{-3}~\langle A~,~A~,~dA~,~A \rangle
\right.\\
&\left.~~~~+~q^{3}~\langle A~,~dA~,~A~,~A \rangle
{}~+~(q+q^{-3})~\langle dA~,~A~,~A~,~A \rangle \right\} \\
&=~[3]^{-1}~\left\{q^{2}~\langle A~,~F~,~F \rangle~+~
\langle F~,~A~,~F \rangle~+~q^{-2}~\langle F~,~F~,~A \rangle \right\} \\
&~~~~-~ig~\{\xi [4][3]\}^{-1}~\left\{(q^{3}+q^{-1})~
\langle A~,~A~,~A~,~F \rangle~+~q~\langle A~,~A~,~F~,~A \rangle
\right.\\
&\left.~~~~+~q^{-1}~\langle A~,~F~,~A~,~A \rangle
{}~+~(q+q^{-3})~\langle F~,~A~,~A~,~A \rangle \right\}
\end{array} \eqno (7.13)$$

Just like those in the classical case [HZ], the gauge fields $F^{J}$ are
invariant under the transformation:
$$A^{J}~\rightarrow~A^{J}~-~\eta^{J},~~~~d~\rightarrow~d~+~\delta
\eqno (7.14) $$

\noindent
In fact,
$$\begin{array}{rl}
F^{J}&\rightarrow~{\cal F}^{J} \\
&=~(d~+~\delta)~(A^{J}~-~\eta^{J}) \\
&~~~~+~\displaystyle {ig \over \lambda}~
\left\{ (A^{0}~-~\eta^{0})(A^{J}~-~\eta^{J})~+~
(A^{J}~-~\eta^{J})(A^{0}~-~\eta^{0}) \right\}\\
&=~F^{J}~+~\left\{\delta A^{J}~-~d\eta^{J}~-~\displaystyle {ig \over
\lambda}~(\eta^{0}A^{J}~+~A^{J}\eta^{0})\right\} \\
&~~~~-~\left\{\delta\eta^{J}~-~\displaystyle {ig \over \lambda}~
\left(\eta^{0}\eta^{J}~+~\eta^{J}\eta^{0}-A^{0}\eta^{J}~-~\eta^{J}A^{0}
\right) \right\} \\
&=~F^{J} \end{array} $$

Now, transforming (7.2) and expanding it by the ghost number, we obtain:
$$\begin{array}{l}
P_{m}~=~(d~+~\delta)~Q_{2m-1}(A-\eta) \\
Q_{2m-1}(A-\eta)~=~\displaystyle \sum_{n=0}^{2m-1}~\omega_{2m-n-1}^{n} \\
P_{m}~=~d \omega_{2m-1}^{0}~+~\displaystyle \sum_{n=0}^{2m-2}~\left\{
\delta \omega_{2m-n-1}^{n}~+~d \omega_{2m-n-2}^{n+1}\right\}~+~
\delta \omega_{0}^{2m-1}
\end{array} \eqno (7.15) $$

\noindent
where the subscripts denote the degrees of form of the quantities,
and the superscripts denote the ghost numbers. In two sides of (7.15)
the quantities with the same degree of form and the same ghost number
should be equal to each other, respectively:
$$\begin{array}{l}
P_{m}~=~d \omega_{2m-1}^{0},~~~~~~\delta \omega_{0}^{2m-1}~=~0\\
\delta \omega_{2m-n-1}^{n}~+~d \omega_{2m-n-2}^{n+1}~=~0,~~~~
n=0, ~1,\cdots,~(2m-2) \end{array} \eqno (7.16) $$

\noindent
For example, for $m=2$ we have:
$$Q_{3}(A-\eta)~=~ \langle A-\eta~,~F \rangle~-~\{\xi [3]\}^{-1}
{}~\langle A-\eta~,~A-\eta~,~A-\eta\rangle $$

\noindent
Simplifying them by the formulas given in Section 6, we obtain:
$$\begin{array}{rl}
\omega_{3}^{0}&=~Q_{3}(A)~=~\langle A~,~F\rangle
{}~-~\{\xi [3]\}^{-1}~\langle A~,~A~,~A \rangle \\
&=~\langle A~,~dA \rangle
{}~+~\displaystyle {[4] \over \xi [3][2]}~\langle A~,~A~,~A \rangle \\
\omega_{2}^{1}&=~-~\langle \eta~,~dA \rangle ~-~\{ig/\lambda \}~
\langle \eta~,~(A^{0}A+AA^{0}) \rangle \\
&~~~~+~\{\xi [3]\}^{-1}~\left\{\langle \eta~,~A~,~A \rangle~+~
\langle A~,~\eta~,~A \rangle~+~\langle A~,~A~,~\eta \rangle
\right\} \\
&=~-~\langle \eta~,~dA\rangle \\
\omega_{1}^{2}&=~-~\{\xi [3]\}^{-1}~\left\{
\langle \eta~,~\eta~,~A \rangle~+~
\langle\eta~,~A~,~\eta \rangle~+~\langle A~,~\eta~,~\eta \rangle
\right\} \\
&=~-~\xi^{-1}~\langle \eta~,~A~,~\eta \rangle\\
\omega_{0}^{3}&=~\{\xi [3]\}^{-1}~\langle \eta~,~\eta~,~\eta \rangle
\end{array} \eqno (7.17) $$

\noindent
It is easy to check by the formulas in Section 6 that (7.17)
satisfies (7.16).

\vspace{10mm}
\noindent
{\bf 8. $q$-deformed Lagrangian and Yang-Mills equation}

In the present paper the spacetime is the ordinary commutative
Minkowski spacetime. Explicitly writing down the spacetime indices,
we have:
$$A^{J}~=~A^{J}_{\mu}~dx^{\mu},~~~~
F^{J}~=~\displaystyle {1\over 2}~F^{J}_{\mu \nu}~dx^{\mu}
\wedge dx^{\nu} \eqno (8.1) $$

\noindent
It is well known that the metric $g^{\mu \nu}$ in the Minkowski spacetime
can change the covariant index to contravariant index, or vice versa.

Now, the $q$-deformed Lagrangian that is covariant both in the Lorentz
transformation and the $q$-gauge transformation is:
$${\cal L}~=~-~\displaystyle {1\over 4}~\langle F_{\mu \nu}~,~
F^{\mu \nu} \rangle ~=~-~\displaystyle {1\over 4}~(F^{I})_{\mu \nu}~
(F^{J})^{\mu \nu}~g_{IJ}  \eqno (8.2) $$

\noindent
We have known that the components of the identity and the adjoint
representations are separated in the $q$-deformed Chern class, and
obviously in the $q$-deformed Lagrangian. Here we only discuss
the $q$-deformed Lagrangian constructed by the adjoint components.

The $q$-deformed Yang-Mills equation is just the $q$-deformed
Lagrangian equation:
$$\begin{array}{l}
\partial_{\mu}~(F^{J})^{\mu \nu}~\left(g_{KJ}+g_{JK}\right) \\
{}~~~~=~(\lambda^{2}+2)^{-1}~\left({\bf C}_{RK}^{~I}-{\bf C}_{KR}^{~I}\right)~
\left\{g_{IJ}(A^{R})_{\mu}(F^{J})^{\mu \nu}~+~
g_{JI}(F^{J})^{\mu \nu}(A^{R})_{\mu}\right\} \end{array} \eqno (8.3) $$

\vspace{1.0cm}
{\bf Acknowledgments}. This work was supported by the National
Natural Science Foundation of China and Grant No. LWTZ-1298 of
Chinese Academy of Sciences.

\vspace{10mm}
\noindent
{\bf References}

\vspace{5mm}
\noindent
[AC] \parbox[t]{14.0cm}{P. Aschieri, L. Castellani, An introduction
to noncommutative differential geometry on quantum groups, {\it Inter.
J. Mod. Phys}. {\bf A8}(1993)1667-1706.}

\vspace{2mm}
\noindent
[AV] \parbox[t]{14.0cm}{I. Ya. Aref'eva and I. V. Volovich, Quantum
group chiral fields and differential Yang-Baxter equations, {\it Phys.
Lett}. {\bf B264}(1991)62-68.}

\vspace{2mm}
\noindent
[Bax] \parbox[t]{14.0cm}{R. J. Baxter, Partition function of the
eight-vertex lattice model, {\it Ann. Phys}. {\bf 70}(1972)193-228.}

\vspace{2mm}
\noindent
[Ber] \parbox[t]{14.0cm}{D. Bernard, Quantum Lie Algebras and Differential
Calculus on Quantum Groups -- Proc. 1990 Yukawa Int. Seminar (kyoto);
{\it Phys. Lett}. {\bf B260}(1991)389.}

\vspace{2mm}
\noindent
[Cas] \parbox[t]{14.0cm}{L. Castellani, $U_{q}(N)$ Gauge Theories,
Preprint, DFTT-74/92, 1992.}

\vspace{2mm}
\noindent
[Con] \parbox[t]{14.0cm}{Alain Connes, Non-commutative differential
geometry, {\it Publ. Math. IHES}. {\bf 62}(1986)41-144.}

\vspace{2mm}
\noindent
[CSWW] \parbox[t]{13.5cm}{U. Carow-Watamura, M. Schlieker, S. Watamura and
W. Weich, Bicovariant differential calculus on quantum groups
$SU_{q}(N)$ and $SO_{q}(N)$, {\it Commun. Math. Phys}.
{\bf 142}(1991)605-641.}

\vspace{2mm}
\noindent
[CW] \parbox[t]{14.0cm}{U. Carow-Watamura and S. Watamura, Complex
quantum group, dual algebra and bicovariant differential calculus,
{\it Commun. Math. Phys}. {\bf 151}(1993)487-514.}

\vspace{2mm}
\noindent
[FP] \parbox[t]{14.0cm}{L. D. Faddeev, P. N. Pyatov, The differential
calculus on quantum linear groups, HEP-TH/9402070.}

\vspace{2mm}
\noindent
[FRT] \parbox[t]{14.0cm}{L. D. Faddeev, N. Y. Reshetikhin and L. A.
Takhtajan, Quantization of Lie groups and Lie algebras, in {\it Algebraic
analysis}, Academic Press, 129, 1988.}

\vspace{2mm}
\noindent
[GJ] \parbox[t]{14.0cm}{D. Gross and R. Jackiw, Effect of anomalies
on quasi-renormalizable theories, {\it Phys. Rev}.
{\bf D6}(1972)477-493.}

\vspace{2mm}
\noindent
[GR] \parbox[t]{14.0cm}{G. Gasper and M. Rahman, Basic Hypergeometric
Series, in {\it Encyclopedia of Mathematics and its Applications 35},
Cambridge, 1990.}

\vspace{2mm}
\noindent
[HHM] \parbox[t]{14.0cm}{Bo-Yu Hou, Bo-Yuan Hou and Zhong-Qi Ma,
$q$-deformed Chern class, Chern-Simons and Cocycle hierarchy,
preprint BIHEP-TH-94-9, HEP-TH/9403153.}

\vspace{2mm}
\noindent
[HZ] \parbox[t]{14.0cm}{Bo-Yu Hou and Yao-Zhong Zhang, Cohomology
in connection space, family index theorem, and abelian gauge
structure, {\it J. Math. Phys}. {\bf 28}(1987)1709-1715.}

\vspace{2mm}
\noindent
[Isa] \parbox[t]{14.0cm}{A. P. Isaev, Quantum group covariant
noncommutative geometry, Dubna Preprint JINR E2-94-38, HEP-TH/9402060.}

\vspace{2mm}
\noindent
[IP] \parbox[t]{14.0cm}{A. P. Isaev and Z. Popowicz, $q$-trace
for the quantum groups and $q$-deformed Yang-Mills theory, Wroclaw
preprint ITP UWr 786/91.}

\vspace{2mm}
\noindent
[Jackiw] \parbox[t]{13.5cm}{R. Jackiw, New directions for topological
research on gauge theories, in {\it Symposium on Anomalies, Geometry,
Topology}, P.156, Eds. W. A. Bardeen and A. R. White, World Scientific,
Singapore, 1985.}

\vspace{2mm}
\noindent
[Jackson] \parbox[t]{13.5cm}{F. H. Jackson, {\it Quart. J. Pure Appl. Math}.
{\bf 41}(1910)193.}

\vspace{2mm}
\noindent
[Jur] \parbox[t]{14.0cm}{B. Jur\~{c}o, Differential calculus on quantized
simple Lie groups, {\it Lett. Math. Phys}. {\bf 22}(1991)177-186.}

\vspace{2mm}
\noindent
[KR] \parbox[t]{14.0cm}{A. N. Kirillov and N. Y. Reshetikhin,
$q$-Weyl group and a multiplicative formula for universal $R$-matrices,
HUTMP 90/B261.}

\vspace{2mm}
\noindent
[Ma] \parbox[t]{14.0cm}{Zhong-Qi Ma, {\it Yang-Baxter Equation and
Quantum Enveloping algebras}, World Scientific, Singapore, 1993.}

\vspace{2mm}
\noindent
[Manin] \parbox[t]{13.5cm}{Yu. I. Manin, Quantum Groups and
Non-Commutative Geometry, Montreal Univ. preprint, CRM-1561, 1988.}

\vspace{2mm}
\noindent
[MSZ] \parbox[t]{14.0cm}{ J. Ma\~{n}es, R. Stora and B. Zumino,
Algebraic study of chiral anomaly,
{\it Commun. Math. Phys}. {\bf 102}(1985)157-174.}

\vspace{2mm}
\noindent
[Res] \parbox[t]{14.0cm}{N. Y. Reshetikhin, Quantum universal
enveloping algebras, the Yang-Baxter equation and invariant
of links I, Preprint, LOMI, E-4-87, 1987.}

\vspace{2mm}
\noindent
[SW] \parbox[t]{14.0cm}{X. D. Sun and S. K. Wang, Bicovariant
differential calculus on quantum group $GL_{q}(n)$, Worldlab-Beijing
preprint CCAST-92-04, 1992.}

\vspace{2mm}
\noindent
[SWZ] \parbox[t]{14.0cm}{P. Schupp, P. Watts and B. Zumino,
Differential geometry on linear quantum groups, preprint,
LBL-32314, UCB-PTH-92/13.}

\vspace{2mm}
\noindent
[Wat] \parbox[t]{14.0cm}{S. Watamura, Quantum deformation of BRST algebra,
{\it Commun. Math. Phys}. {\bf 158}(1993)67-92.}

\vspace{2mm}
\noindent
[Wor1] \parbox[t]{14.0cm}{S. L. Woronowicz, Compact matrix pseudogroups,
{\it Commun. Math. Phys}. {\bf 111}(1987)613-665.}

\vspace{2mm}
\noindent
[Wor2] \parbox[t]{14.0cm}{S. L. Woronowicz, Differential calculus
on compact matrix pseudogroups (quantum groups), {\it Commun. Math. Phys}.
{\bf 122}(1989)125-170.}

\vspace{2mm}
\noindent
[WZ] \parbox[t]{14.0cm}{J. Wess and B. Zumino, Covariant differential
calculus on the quantum hyperplane, {\it Nucl. Phys}. (Proc. Suppl.)
{\bf B18}(1990)302-312. }

\vspace{2mm}
\noindent
[Yang] \parbox[t]{14.0cm}{C. N. Yang, Some exact results for the
many-body problem in one dimension with repulsive delta-function,
{\it Phys. Rev. Lett}. {\bf 19}(1967)1312-1315. }

\vspace{2mm}
\noindent
[Zum1] \parbox[t]{14.0cm}{B. Zumino, Chiral Anomalies and
Differential Geometry, Lectures given at Les Houches, Preprint,
LBL-16747, 1983.}

\vspace{2mm}
\noindent
[Zum2] \parbox[t]{14.0cm}{B. Zumino, Introduction to the differential
geometry of quantum groups, LBL-31432 and UCB-PTH-62-91, Notes of a
plenary talk given at the 10th IAMP Conf., Leipzig, 1991.}

\end{document}